\documentclass{elsart}

\usepackage{graphicx}
\usepackage{amssymb}

\begin{document}

\begin{frontmatter}



% editor advice: concise and informative
%\title{Timing analysis of the MAGIC telescope data}
\title{Improving the performance of the single-dish Cherenkov
  telescope MAGIC through the use of signal timing}


\author[a]{E.~Aliu},
\author[b]{H.~Anderhub},
\author[c]{L.~A.~Antonelli},
\author[d]{P.~Antoranz},
\author[e]{M.~Backes},
\author[f]{C.~Baixeras},
\author[d]{J.~A.~Barrio},
\author[g]{H.~Bartko},
\author[h]{D.~Bastieri},
\author[e]{J.~K.~Becker},
% ...
\address[a] {IFAE, Edifici Cn., Campus UAB, E-08193 Bellaterra, Spain}
\address[b] {ETH Zurich, CH-8093 Switzerland}
\address[c] {INAF National Institute for Astrophysics, I-00136 Rome, Italy}
\address[d] {Universidad Complutense, E-28040 Madrid, Spain}
\address[e] {Technische Universit\"at Dortmund, D-44221 Dortmund, Germany}
\address[f] {Universitat Aut\`onoma de Barcelona, E-08193 Bellaterra, Spain}
\address[g] {Max-Planck-Institut f\"ur Physik, D-80805 M\"unchen, Germany}
\address[h] {Universit\`a di Padova and INFN, I-35131 Padova, Italy}
% ...
\corauth[cor1]{Corresponding author.}
%\address{diegot@ifae.es, moralejo@ifae.es}

 \begin{abstract}
% Text of abstract
The Cherenkov light flashes produced by Extensive Air Showers
are very short in time.
%A fast readout, 
A high bandwidth and fast digitizing readout, therefore, can minimize the influence 
of the background from the light of the night sky, and improve the performance in 
Cherenkov telescopes. 
The time structure of the Cherenkov image can further be used in single-dish 
Cherenkov telescopes as an additional parameter to reduce the background 
from unwanted hadronic showers. 
A description of an analysis method which makes
use of the time information and the subsequent improvement on the
performance of the MAGIC telescope (especially after the upgrade with
an ultra fast 2~GSamples/s digitization system in February 2007) will
be presented. The use of timing information in the analysis of the new
MAGIC data reduces the background by a factor two, which in turn
results in an enhancement of about a factor 1.4 of the flux
sensitivity to point-like sources, as tested on observations of the Crab Nebula.
\end{abstract}

\end{frontmatter}

%
%%%%%%%%%%%% INTRODUCTION %%%%%%%%%%%%%
%
\section{Introduction} \label{intro}
Imaging Atmospheric Cherenkov Telescopes (IACTs) collect the Cherenkov
light  from Extensive Air Showers (EAS) to form an image. 
The morphology of the shower image \cite{hillas_parameters} is
used to recognize the few $\gamma$-ray initiated showers among the
much more numerous hadronic showers initiated by cosmic ray nuclei.
This standard approach only exploits the knowledge of the spatial distribution of the Cherenkov photons in the camera plane, but further informations regarding the shower development are in principle available in the photon arrival times \cite{hillas_time}.  
The possibility of using effectively the timing information to improve the performance 
of IACTs has been explored in earlier works.
The HEGRA collaboration measured on their data a time gradient along the 
major axis in the Cherenkov images \cite{TimingHEGRA1999}.
They suggested that this information may be useful to estimate the
distance to the shower core and the shower direction in the case of a
single Cherenkov telescope, but of limited use in an array of IACTs,
where a stereoscopic view of the shower is available.
A recent MC study \cite{delacalle} suggests that the use of the
time profile of Cherenkov images may lead to important background
rejection improvements in future Cherenkov instruments (even if, according to \cite{veritas}, 
pioneering tests on real data data led only to marginal improvements). 
A different approach to exploit the time information is proposed by the authors
of \cite{2006APh....25..342M}, capitalizing on the different characteristic time
spread of the images of gamma-initiated air showers as compared to hadronic 
showers or images from distant single muons.

The MAGIC (Major Atmospheric Gamma Imaging Cherenkov) telescope is a
single-dish Cherenkov telescope, designed for the detection of VHE
gamma rays in the $\sim50$~GeV to $\sim10$~TeV band
\cite{2005ICRC....5..359C}. Its camera is composed of 577 pixels
equipped with high quantum efficiency photomultiplier tubes (PMTs).
In the first years of operation of MAGIC, the PMT signals were digitized 
with 300~MSamples/s Flash Analogic to Digital Converters (FADCs).
In February 2007 the data acquisition of the MAGIC telescope was
upgraded with ultra-fast FADCs capable to  digitize at 2~GSamples/s
\cite{MirzoyanMultiplexer,2007ICRCGoebel}. 
The implementation of a faster readout might lead to an improvement in the
telescope performance for two reasons: a reduction in the amount of
NSB (Night Sky Background) light integrated with the real signal, and
an improvement in the reconstruction of the timing characteristics of
the recorded images.
The main aim of this work is to establish whether the timing
information is useful in the analysis of single-dish
IACT data. In the following we will present an analysis method which
makes use of signal timing, and compare its performance to that of the
standard MAGIC analysis used up to now.
\par
The timing analysis proposed here is composed of two different parts. 
The first is the use of the time information to enhance the efficiency and 
to lower the threshold of the image cleaning procedure, thanks to the introduction 
of time constraints. 
The second is the use of additional time-related image parameters in 
the algorithms for the suppression of the isotropic background of hadron-initiated 
showers. 
Although the possibility of using timing to improve the IACT technique
was suggested a long time ago, this is, to our knowledge, the first
time in which it has been successfully applied to real data.

%
%%%%%%%%% ANALYSIS METHOD %%%%%%%%%%%%%
%
\section{Analysis method} \label{meth}
When an atmospheric shower triggers the MAGIC telescope, the
information of all camera pixels is stored by the Data AcQuisition
(DAQ) system. 
This information consists mainly of the digitized pulse
of the PMT corresponding to each pixel in time slices of 0.5 nanoseconds. 
From the digital information of the pulse it is possible, through the so-called 
signal extractor routine, to reconstruct the number of photons that arrived 
at the pixel and their mean arrival time.  
This can be done in several manners. 
For the current MAGIC data (with 2 GS/s sampling), a simple cubic spline 
is built from the FADC readout, and its integral in a range around the highest 
peak provides a measure of the charge recorded by the pixel. 
The arrival time is defined as the position of the rising edge of the pulse
at 50\% of the peak value. 
Before the upgrade of the FADC system, the pulse shape
and duration was dictated by the artificial stretching introduced in
the electronic chain to ensure that the pulse spanned over several
FADC samples (then taken every 3.3 ns). 
For those older data, the {\it digital filter} algorithm \cite{DigitalFilter}, which
makes use of the known pulse shape \cite{signal_extraction}, was used. 
After calibration, the charge (Q) is converted to photo-electrons units (phe). 
Details about the calibration can be found in \cite{GaugPhDthesis}.
\subsection{Image Cleaning} \label{cleaning}
\begin{figure}[!h]
\begin{center}
\includegraphics[width=0.49\textwidth,height=0.42\textwidth]{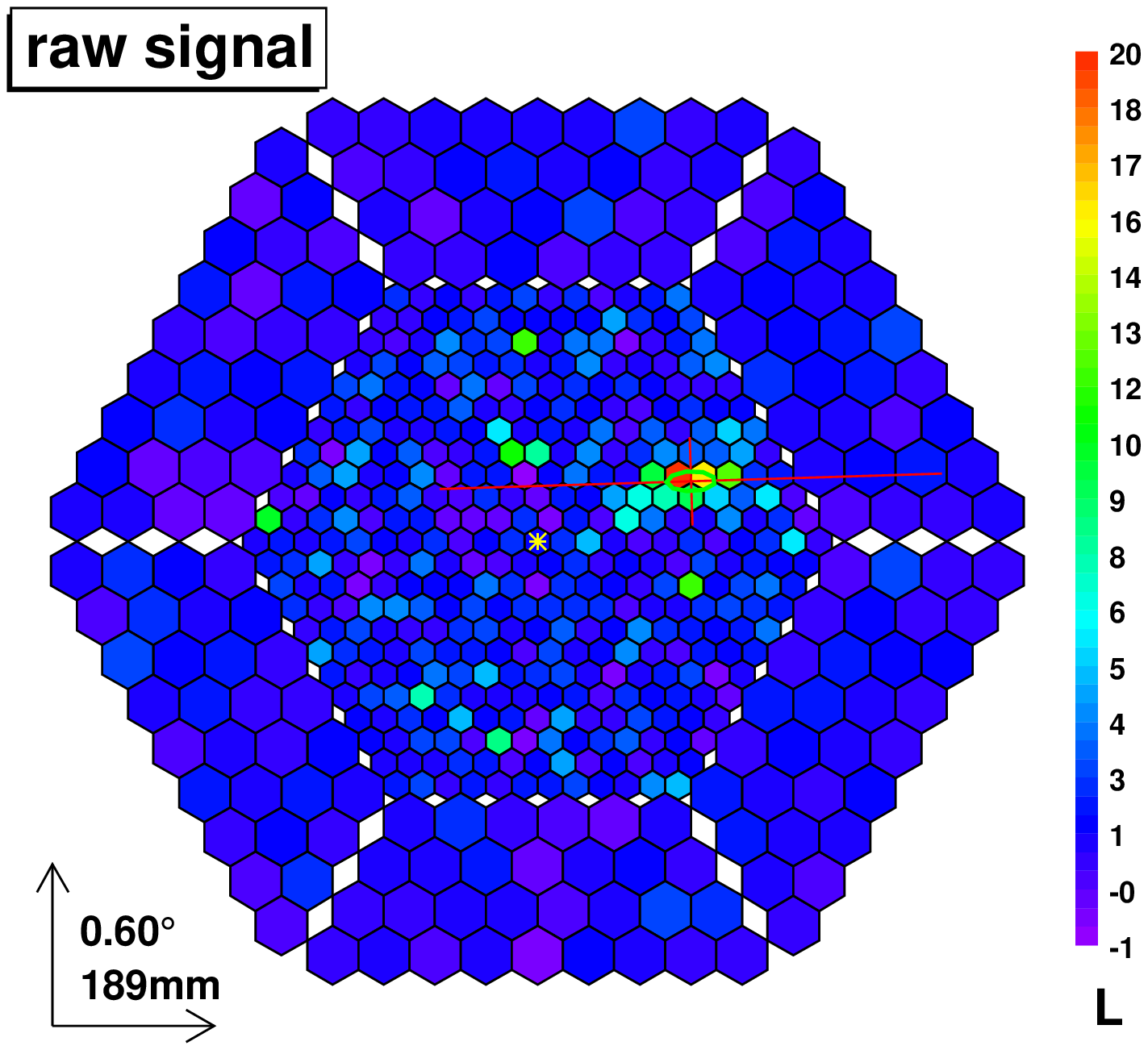}
\includegraphics[width=0.49\textwidth,height=0.42\textwidth]{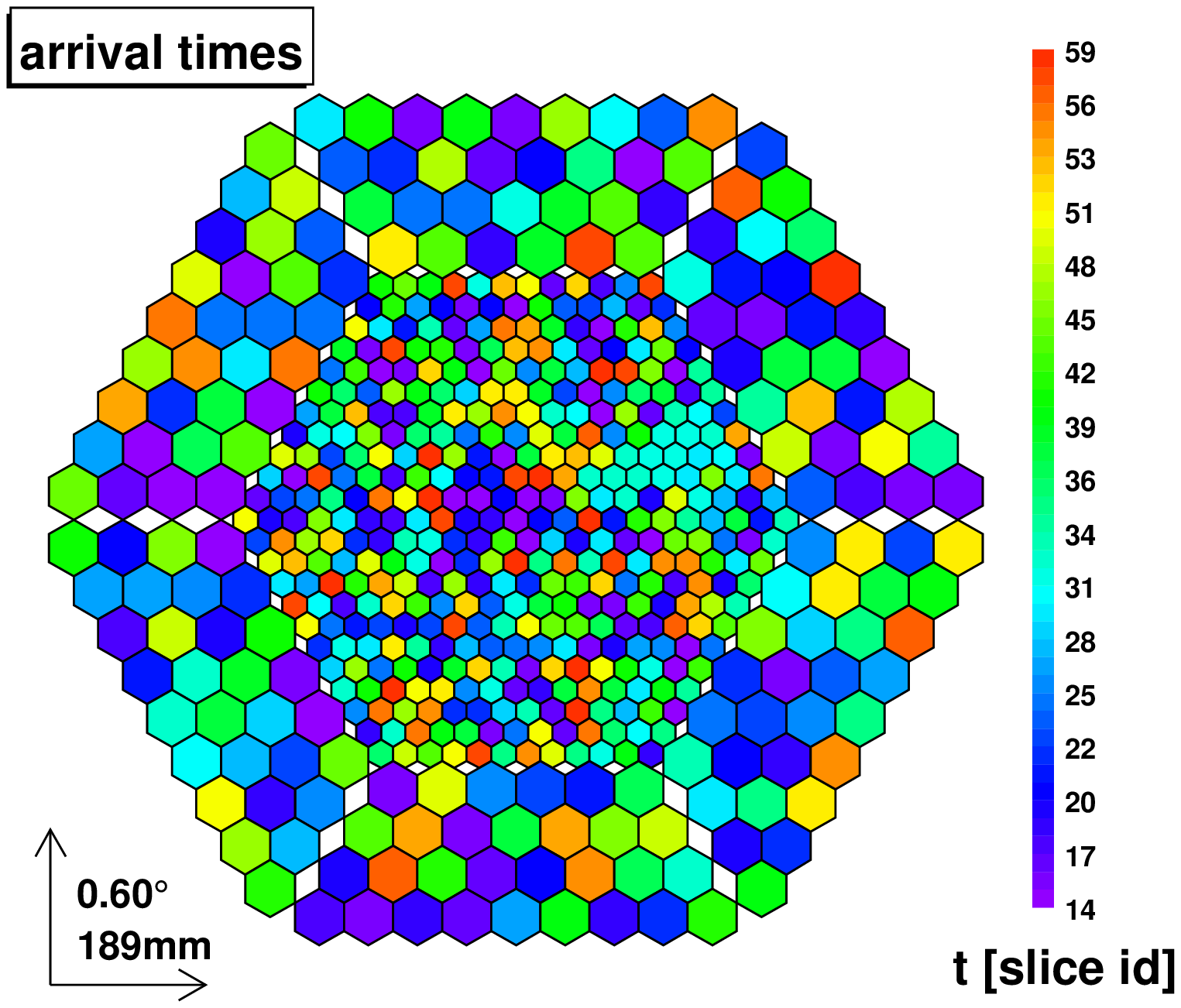}
\includegraphics[width=0.49\textwidth,height=0.42\textwidth]{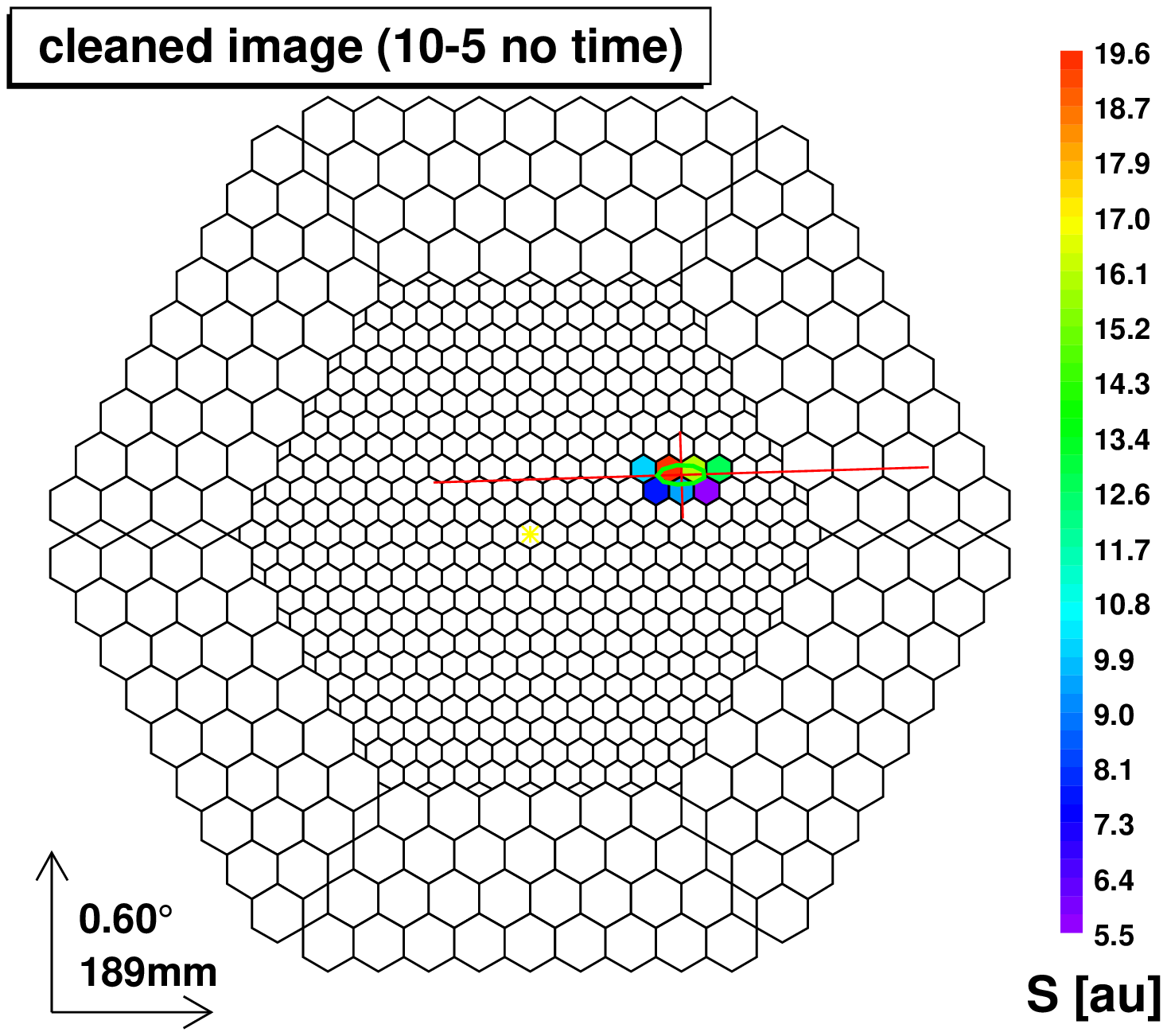}
\includegraphics[width=0.49\textwidth,height=0.42\textwidth]{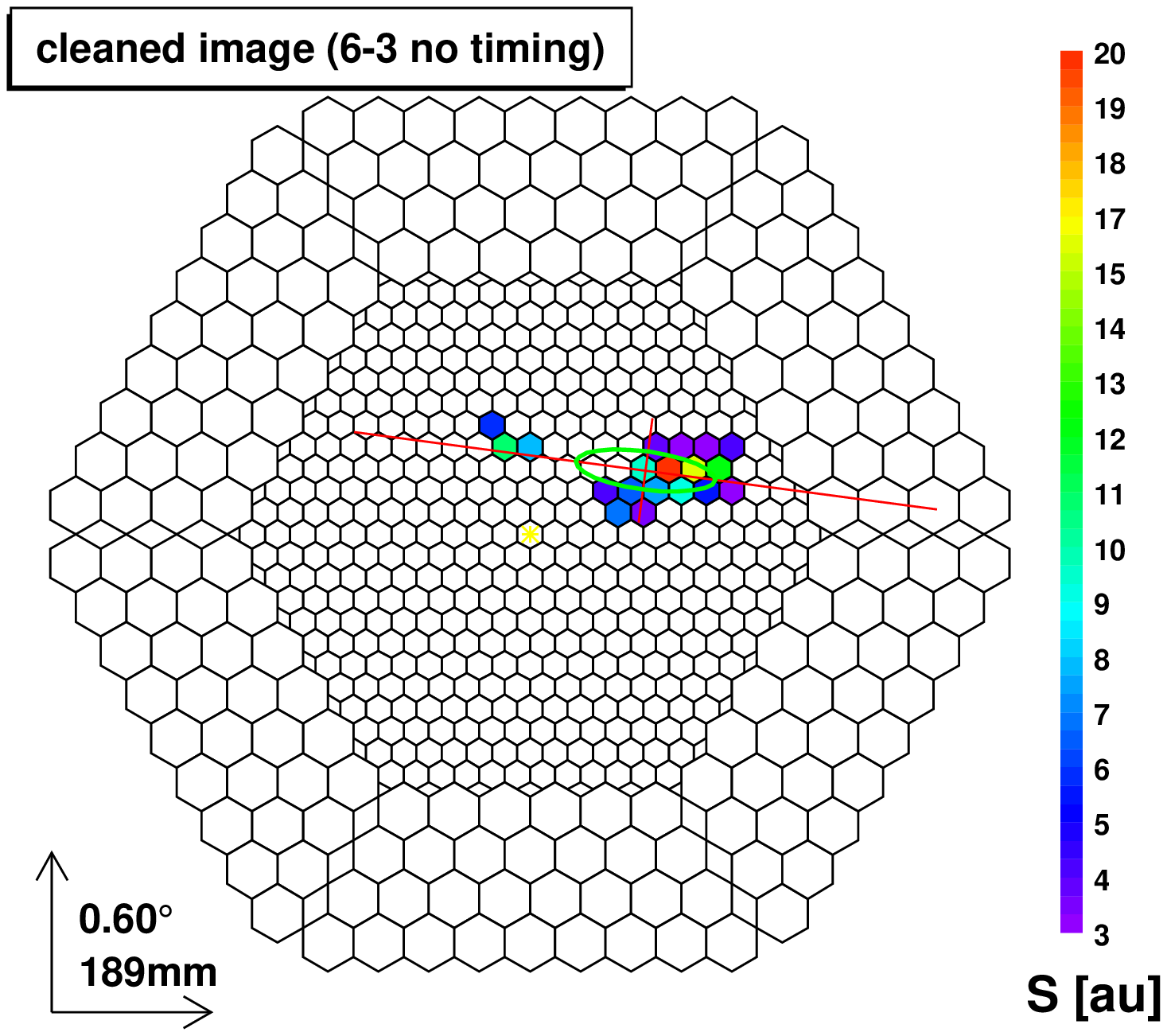}
\includegraphics[width=0.49\textwidth,height=0.42\textwidth]{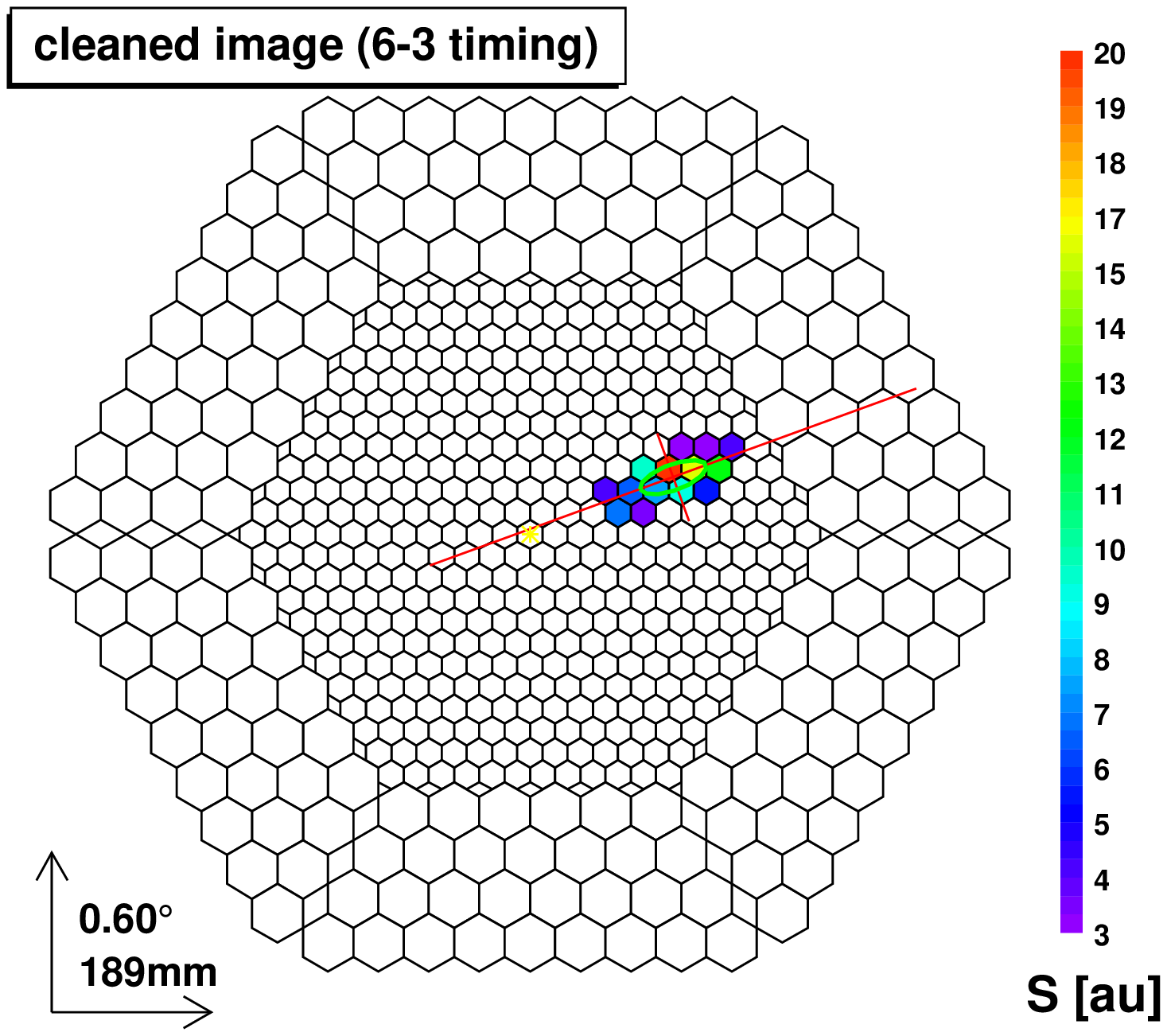}
\end{center}
\caption{Illustrative $\gamma$-event images (Monte~Carlo data, Energy=71~GeV, Impact~Parameter=111~m). 
First row: display of raw recorded data (left) and arrival times information (right).
Second row: comparison of standard cleaning with 10-5 phe minimum charge levels (left) and 6-3 minimum charge levels (right).
Bottom: image obtained with the time image cleaning (6-3 phe minimum charge levels and 4.5~ns and 1.5~ns as time constrains).
The simulated gamma-ray source is located in the center of the camera (yellow star).} \label{imm:event_display}
\end{figure}

The information from the pixels is first used to perform the
\emph{image cleaning}, that aims at identifying which pixels belong
to the shower image. 
In figure \ref{imm:event_display} an example of an event before and 
after the cleaning is shown.
\par
In the MAGIC Analysis and Reconstruction Software (MARS
\cite{WagnerMARS}), different cleaning methods can be chosen by the
user. 
The most commonly used is the \emph{standard - absolute} method. 
The choice may depend on the sky around the source (galactic or extra-galactic) 
or the prevailing atmospheric conditions.
This procedure uses a threshold signal value $q_1$ (a fixed
value in terms of phe) to select the \emph{core pixels},
namely all those with charge above $q_1$ and which have at least
one neighbor fulfilling the same condition\footnote{This additional
  requirement avoids the selection of pixels unrelated to the image
  whose large charge results from an afterpulse in the PMT.}. 
In a second stage, all pixels which have at least one {\it core} neighbor,
and whose charge is above $q_2$ (with $q_2 < q_1$), are included in
the image (these are called \emph{boundary pixels}). 
\par
Relaxing the cleaning levels $q_1$ and $q_2$ results in a larger number 
of pixels per image, and accordingly a lower analysis energy threshold, 
since a minimum number of pixels is needed to proceed with the analysis. 
On the other hand, a low cleaning level increases the probability to
include in the cleaned image a noise pixel (mainly due to
NSB or other unwanted light pollution). 
The inclusion of pixels unrelated to the shower degrades the image parameters 
and worsens the performance of the subsequent analysis.
\par
Together with the signal intensity also an arrival time value is
assigned by the signal extractor to each pixel. These times can be
used to further constrain the selection of core and boundary pixels in
the image cleaning algorithm: Cherenkov flashes are very brief (of the
order of few ns), and NSB photons produce pulses asynchronous 
with respect to the pulses of the shower image. 
A timing coincidence window between the mean arrival time and the single 
pixel arrival time can avoid to confuse NSB signals with real image tails. 
This further constraint allows to relax the cleaning levels $q_1$ and $q_2$, 
lowering in this way the energy threshold. 
The time information has already been used for the image cleaning in
the analysis of the observations of the Crab Nebula with MAGIC \cite{2007CrabMAGIC}. 
However, the algorithm used in that analysis differs from the one we are proposing here and can be found in detail in \cite{NepomukThesis}.\\
The procedure used in this work can be summarized in this way:
\begin{itemize}
\item After selecting the core pixels in the same way as in the
  standard procedure, we reject those whose arrival time is not within
  a time $\Delta t_1$ of the mean arrival time of all core pixels.
\item In the selection of the boundary pixels we add the constraint
  that the time difference between the boundary pixel candidate and
  its neighbor core pixels is smaller than a second fixed time
  constraint $\Delta t_2$.
\end{itemize}
The charge levels of the standard cleaning commonly used in the past
in the analysis of MAGIC data are $q_1 = 10$ phe for the core pixels
and $q_2 = 5$ phe for the boundary pixels. 
For the time-cleaning approach, the charge threshold levels were decreased 
to 6 and 3 phe respectively. 
Concerning the time constraints, the values $\Delta t_1 = 4.5$ ns and 
$\Delta t_2 = 1.5$ ns were selected. 
The choice of these values is supported by a study based on Monte Carlo data 
(see \cite{ICRC_timing} and \cite{tescaro_tesina} for more details), in which we 
have assumed ``dark night'' conditions (and hence the used criteria would not be optimal 
during moon light or twilight observations when the number of noise photons is higher).
The setting of these time constrains resulted also not very critical for choices within 
$\simeq$~1ns respect to the values used here.\\
In figure \ref{imm:event_display} an example event is shown.
The image footprint is visible in the arrival time display (upper right plot) because of
the short duration of the Cherenkov flash, illustrating the validity of the time image 
cleaning approach. 
The arrival times of the signal pixels are distributed within few ns. 
The other pixels have, as expected, a random arrival time distribution.
In the second and third rows of figure \ref{imm:event_display} the same 
event is plotted after applying different cleaning methods.
%
% IMAGE PARAMETERS %%%%%%%%%%%%%%%%%%%%%
%
\subsection{Timing characteristics of the shower images}
\begin{figure}[!t]
\centering
\includegraphics[width=1.0\textwidth]{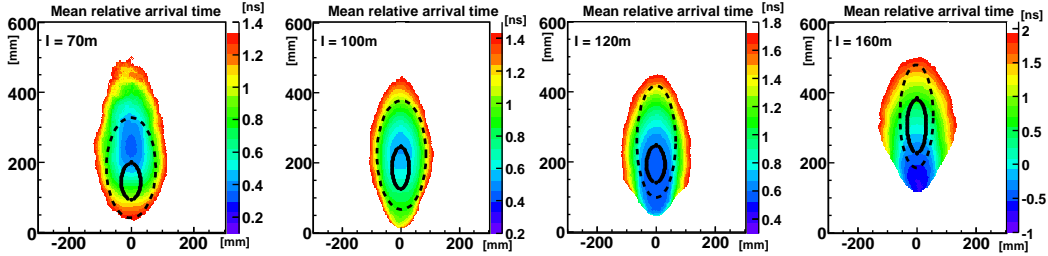}
\caption{Relative arrival time distributions of photons, averaged over a large sample of Monte~Carlo $\gamma$-events (Energy=100~GeV) for several values of the impact parameter.
  The black solid line marks the 50\% of the maximum photon density whereas the dashed line 
  ($\approx$10\% of the maximum) represents qualitatively the border of the image after the cleaning procedure. 
  The time profile of the recorded images changes clearly depending on the IP of the primary shower.
  The source is located in the (0,0) position.
  Plots from \cite{MazinModelAnalysis}.} \label{model_analysis} 
\end{figure}

As previously introduced, Cherenkov images present some timing features, the most 
important of which is a dependency between the timing profile along the major axis 
of the image and the Impact Parameter (IP) of the shower. 
The model proposed in \cite{TimingHEGRA1999} explains well this relationship.
In case of a small impact parameter (IP $\leq 60$~m), the light emitted in the higher
part of the shower (the \emph{shower head}) will arrive delayed with respect
to the light emitted in the lower part of the shower (the \emph{tail}), since the photons 
emitted first travel slower (at a speed $c/n$) than the ultra relativistic particles of the 
shower that produce the photons at lower altitudes.
In case of a larger impact parameter (IP $\geq120$~m), the effect just described is reduced or even inverted, as the arrival time from the tail becomes the sum of  the times spent in the paths of particles and photons, respectively.
In this latter situation, the photons emitted in the lower part of the shower will arrive later than the photons emitted in the upper part.
Events with an intermediate impact parameter show a flat time profile.
These features are well visible in the templates of average Monte Carlo gamma-ray images on the MAGIC camera (figure \ref{model_analysis}), created by the superposition of many events at fixed values of energy and impact parameter. 
These are part of a dedicated MC sample produced for a different study \cite{MazinModelAnalysis} 
on the applicability of the so-called ``model analysis'' (\cite{deNaurois}) to the MAGIC data. 
In these templates, it is possible to recognize the dependency of the timing structure 
with the IP: the arrival time increases from shower head (bottom part of the images) 
to shower tail at large impact parameter, and from tail to head for small impact parameters.
\subsection{Definitions of time-parameters} \label{timepar_description}
In order to exploit the timing characteristics of the showers in the analysis
stage, some time-related image parameters have to be introduced.
A linear fit of the arrival time versus the distance along the major
image axis provides an easy way to characterize the time profile of a
shower image. 
Another useful quantity may be the overall spread of the arrival times of all 
pixels surviving the cleaning. 
Based on these considerations, two new time-related image parameters 
have been introduced:
\begin{itemize} 
\item {\bf \emph{Time Gradient}}: this parameter measures how fast the
  arrival time changes along the major image axis. The pixel
  coordinates are projected onto this axis, reducing the problem to one dimension. 
  Then the arrival time versus the space coordinate along the major axis is fitted to 
  a linear function $t=m \cdot x + q$. 
  The slope $m$ is called in the following \emph{Time Gradient} of the image. 
  The sign of this parameter is positive if the
  arrival time increases as we move away from the location of the
  source on the camera, negative otherwise. It is therefore a
  parameter which depends on the position of the candidate gamma-ray
  source.
%  More complex functions can be used to obtain a better fit, specially for IP<100 m, but since the linear coefficient has a clear physical meaning (time profile orientation of the image, positive negative or flat) and results already effective, the simplest approach is preferred.
\item {\bf \emph{Time RMS}}: the root mean square of the arrival times of all
  pixels belonging to the image after cleaning. It measures the
  spread of the arrival times irrespective of the pixel position in the camera.
  This parameter has been suggested as a possible background
  discriminator in \cite{2006APh....25..342M}. It must be noted that due to the 
  time structure of the events, this parameter is correlated with
  the \emph{Time Gradient}.
\end{itemize}
\begin{figure}[ht]
   \centering
%\includegraphics[width=0.47\textwidth]{./fig3-1.eps} 
%\hspace*{0.5cm}
%\includegraphics[width=0.47\textwidth]{./fig3-2.eps}
\includegraphics[width=0.48\textwidth]{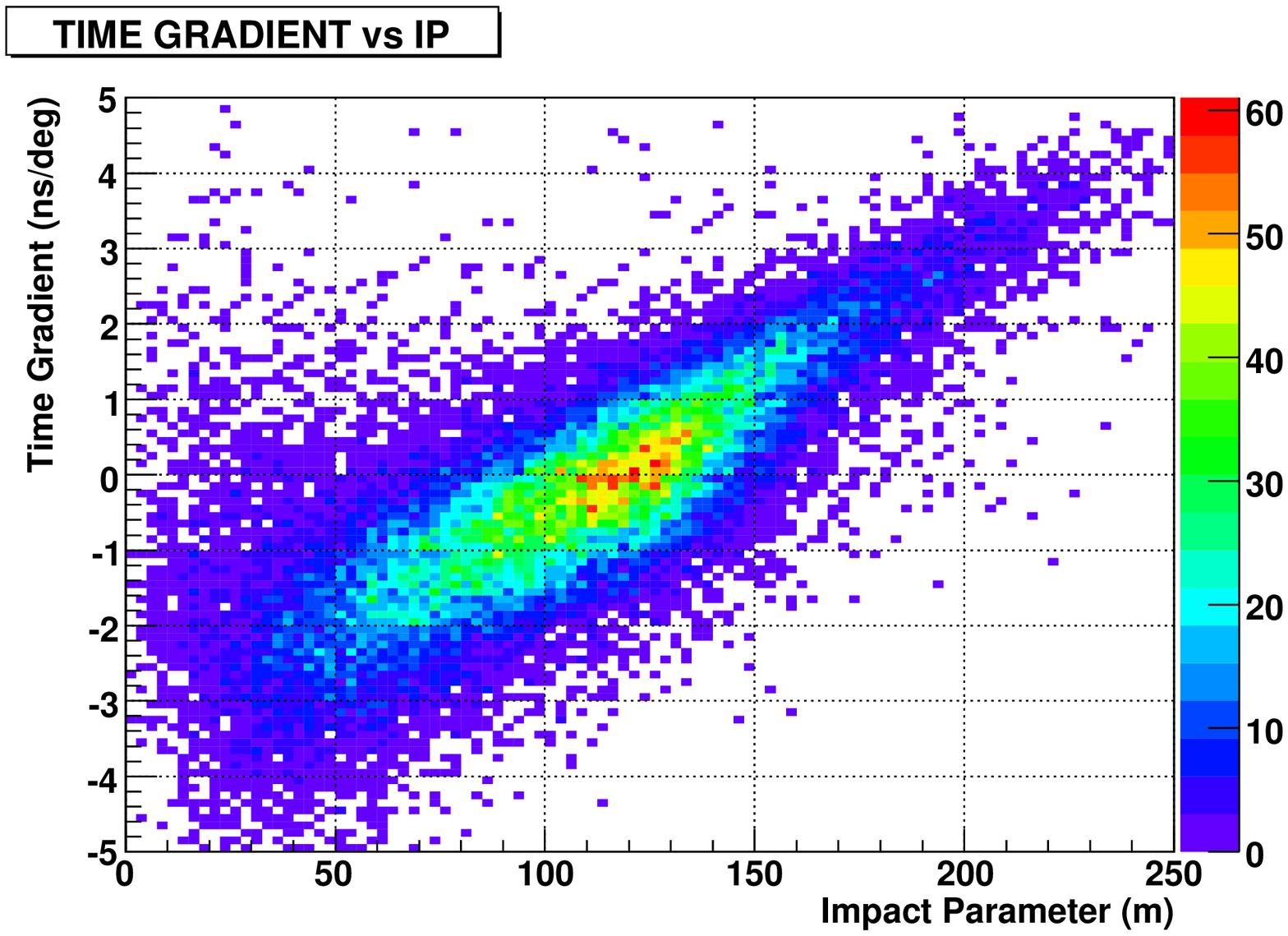} 
\includegraphics[width=0.48\textwidth]{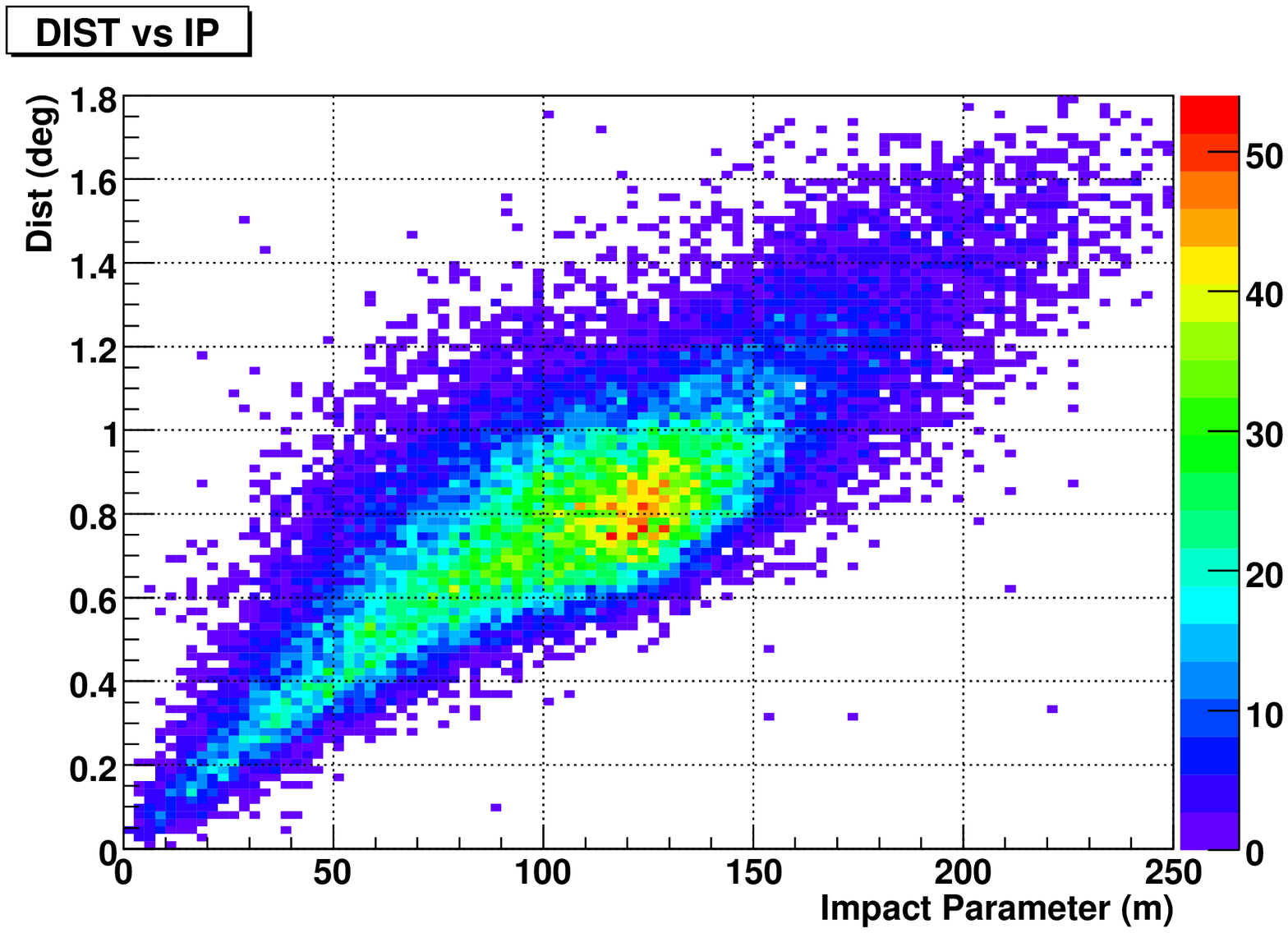}
   \caption{Scatter plot of \emph{Time~Gradient} vs IP (left) and scatter
     plot of \emph{Dist} vs IP (right) for Monte Carlo $\gamma$-events. 
     A cut \emph{Size} $>200$ phe has been applied. 
     Note that the impact parameter
     correlates better with \emph{Time Gradient} for distant showers,
     whereas the correlation is better with \emph{Dist} for nearby
     showers. The information provided by the two image parameters is
     hence complementary.}
   \label{correlations1}
\end{figure}
\par
For the geometrical reasons explained in the previous section, the
\emph{Time Gradient} is well correlated with the impact parameter, as can 
be seen in the left panel of figure \ref{correlations1}.
On the other hand, the classical \emph{Dist} parameter, which is the
angular distance from the image center of gravity to the source
location on the camera, is also correlated to the impact parameter for
gamma rays coming from a point-like source: as we increase
the impact parameter, the image gets longer and moves away from the
source, as we observe it at an increasingly larger
angle\footnote{Fluctuations in the shower development make that, even
  for a fixed energy, the altitude at which it develops
  changes from event to event, which blurs the correlation of \emph{Dist}
  and the impact parameter.}. 
A consequence of this is that \emph{Time Gradient} is correlated with 
\emph{Dist} for gamma-ray images from a point-like source (see fig. \ref{time_gradient}), 
whereas no such correlation exists for hadron images, since hadron 
showers are distributed isotropically, and therefore no strong correlation 
of \emph{Dist} and the impact parameter is expected for them. 
Already from this, one can expect some improvement
in the background discrimination through the use of the \emph{Time Gradient} in the analysis. 
\par
Another way of looking at this is the following: the shower direction
is not well determined by a single IACT. 
When observing a point-like source, all gamma-ray images will be pointing 
(within $\simeq10^\circ$) towards the source location on the camera, but so will 
many background cosmic ray showers whose axes are coplanar with the line pointing
from the mirror dish center toward the source. 
The bare shower shapes allow to eliminate some of those, but the timing profile
provides additional independent information to recognize the
gamma rays (the images with ``consistent'' values of \emph{Dist} and the \emph{Time 
Gradient}) and reject the background, and is therefore expected to improve the 
performance of the analysis. 
\par
Note that in stereoscopic IACT systems the determination of the shower direction 
and the impact parameter is obtained by the intersection of the multiple shower images,
and thus the information that could be provided by the timing is redundant\footnote{In the special case of a two telescopes stereo system, the impact parameter can still be poorly determined for some degenerate events.}.
Therefore, the results obtained in this study should not be extrapolated to the case of stereo observations. 
\par
Regarding the \emph{Time RMS}, it has been suggested in \cite{2006APh....25..342M} 
that it may be of help in identifying triggers produced by single, large impact 
parameter muons (whose images may otherwise be gamma-like), as well as 
other hadron-initiated showers (since their \emph{Time RMS} distribution has, 
with respect to that of gamma rays, a longer tail towards large values).
\subsection{Role of the Monte Carlo simulation} \label{MCSimulation}
Making sure that the Monte Carlo reproduces the features of the real
data is very important to perform a good background rejection
and energy estimation. 
In the MAGIC analysis, both tasks rely heavily on the MC simulated events. 
The MC is also crucial when the gamma-ray
flux of a source is computed, since the estimation of the collection
area is done using a Monte Carlo ``test'' sample. 
Therefore, the detector simulation has been updated to reproduce the digitization  
features of the new 2 GS/s digitization system: beyond the higher
digitization speed, also the level of electronic noise and the
overall precision of the time determination have been adjusted, taking
into account the entire electronics chain.
The time resolution can be estimated from the calibration events 
(light pulses of $\sim$2~ns duration), looking at the distribution
of the arrival time difference between any two camera pixels. 
The RMS of the distribution is 550~ps. 
This correspond to a time resolution for a single pixel of $550/\sqrt{2} =  390$~ps 
(see \cite{2007ICRCGoebel} for details). 
Actually, Cherenkov pulses are generally faster than the calibration pulses, and hence,
for a pulse of comparable amplitude, showers signals have a better resolution.
\par
\begin{figure}[!t]
\begin{center}
\includegraphics[width=1.0\textwidth,height=0.28\linewidth]{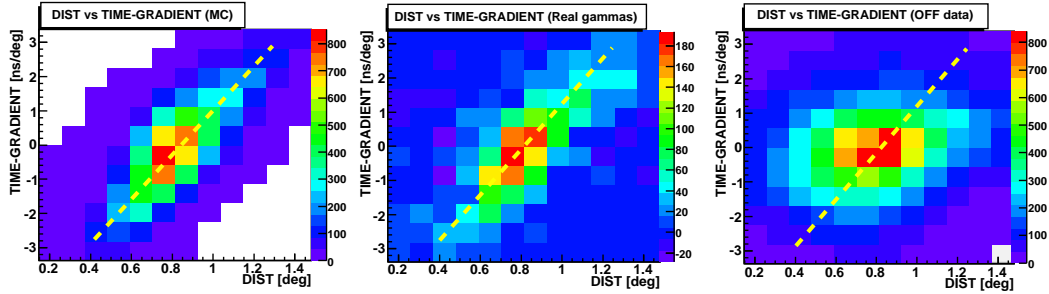}
\end{center}
\caption{\emph{Time Gradient} versus \emph{Dist} parameter correlation for
  MC $\gamma$-rays (left), real $\gamma$-rays (center) and background
  (right) data. 
  } \label{time_gradient} 
\end{figure}

A demonstration of the \emph{Time~Gradient} - \emph{Dist} correlation described in section
\ref{timepar_description} can be seen in figure \ref{time_gradient}. 
The left plot of the figure is made with pure
$\gamma$-MC events while the central panel displays the difference 
between \emph{on}-source and \emph{off}-source distributions (from a Crab sample
described later), and therefore shows the distribution of the
gamma ray excess. 
A correlation \emph{Time~Gradient} - \emph{Dist} is present in both cases. 
Such correlation is almost completely suppressed for hadron images (even after a cut in the \emph{Alpha} parameter), as shown in the plot on the right.
%
%%%%%%%%%%%%%%%%% EXPERIMENTAL RESULTS %%%%%%%%%%%%
%
\section{Experimental results}\label{results}

MAGIC observations are performed mainly in two modes: \emph{on-off} and
\emph{wobble}. 
In the former, the telescope points directly at the source to obtain the 
\emph{on}-source data whereas the \emph{off} data, used to estimate the background, 
are taken by pointing at a region of the sky where no signal is expected. 
The wobble mode eliminates the need for taking dedicated \emph{off}
runs \cite{Fomin}. 
The telescope is not aimed directly at the source, but slightly off ($0.4^\circ$ away). 
In this way, the source does not occupy a privileged position in the camera, 
and the background can be estimated by re-doing the analysis
with respect to points on the camera (``false-sources'') which are expected to be
equivalent to the source location (for instance the point symmetric to the source 
with respect to the camera center). 
The ``wobbling'' consists in changing the telescope pointing
every twenty minutes between two symmetric sky directions around the
source, which is an additional guarantee of the equivalence of source
and false-source against effects like inhomogeneities in the camera
response or the dependence of acceptance with the zenith angle. 
For the studies presented in this paper, the background has always been
estimated from one single false-source, located opposite to the source w.r.t. 
the camera center.
The main disadvantage of the wobble method is a small reduction of the
trigger efficiency leading to a reduction of $\simeq$~15-20\% in the nominal 
flux sensitivity, since the trigger area is limited to $\simeq 1^\circ$ around the 
camera center.
\par
The data sample chosen for this study consist of 5.7~h of Crab Nebula
observations performed in wobble mode during the nights of the
$7^{th}$, $9^{th}$, $15^{th}$ and $17^{th}$ of February 2007 (soon
after the installation of the new MUX FADCs readout) at a zenith angles
smaller than $30^\circ$. Weather conditions were good during all the nights
considered.
\subsection{Analysis comparison strategy} \label{analyses}
In order to compare the sensitivity with and without the help of the
timing information, three different analyses of the above mentioned
Crab Nebula data sample were performed: 
\begin{enumerate}
\item The standard analysis commonly performed on the MAGIC
  data before the upgrade of the DAQ. The image cleaning levels were
  10 and 5~phe (see section \ref{cleaning}), and no time information
  was used. The standard image parameters (\emph{Size}, \emph{Width}, \emph{Length}, \emph{Dist},
  \emph{Conc} and the third moment along the major axis, dubbed
  \emph{M3long}\footnote{This measures the image asymmetry along its major
    axis. It is a source-dependent parameter, since its sign
    is referred to the source position on the camera. The sign is
    defined such that it is positive when the shower head is closer to
    the source than the shower tail, as is the case for properly
    reconstructed gamma rays.}) were used to perform the $\gamma$/h
  separation. This is the reference analysis for the comparison.

\item An analysis using 6-3~phe as cleaning levels, with the time
  constraints described in section \ref{cleaning}.
  The same standard parameters of analysis 1 were used for $\gamma$/h
  separation. This analysis is meant to evaluate the effect of the
  time cleaning. 

\item The same 6-3~phe time cleaning of analysis 2 is used. In this
  analysis, in addition to the standard image parameters, the \emph{Time RMS}
  and the \emph{Time Gradient} image parameters (see section
  \ref{timepar_description}) were used as input  for the background 
  rejection. This analysis is meant to evaluate the analysis
  improvement due to the timing parameters (used together with the
  time cleaning).
\end{enumerate}
In all cases the image parameters were the input to the Random Forest
(RF) event classification algorithm \cite{RFreference}, which was used
to perform the $\gamma$/h separation task. 
The training samples for the construction of the RF are a MC gamma sample,
and a sample of real \emph{off} data to represent the background. 
When applied to the data, the RF tags each event with a single value called 
\emph{Hadronness} (ranging from 0 to 1) which is a measure, based on the
image parameters, of the likelihood that the event is a background
event. 
\par
The sum of the signals (in phe) of the two pixels with highest signal (\emph{Size-2}) 
was used as parameter to select event samples of different energies. 
Like the classical event \emph{Size}, \emph{Size-2} is correlated with energy, 
but unlike \emph{Size}, it is very weakly dependent on the cleaning
levels\footnote{Considering two different image cleanings and applying
  the same \emph{Size-2} cut, the two data samples obtained will contain
  essentially the same events, differing only in the events that
  survive just one of the cleanings.}.
If we had chosen the total \emph{Size} to define the samples, we would
have faced the problem that they would correspond to different
energies in the three analyses, therefore making the interpretation of
the results more difficult.
\par
\begin{figure}[!h]
\begin{center}
\includegraphics[width=0.42\textwidth,height=0.38\textwidth]{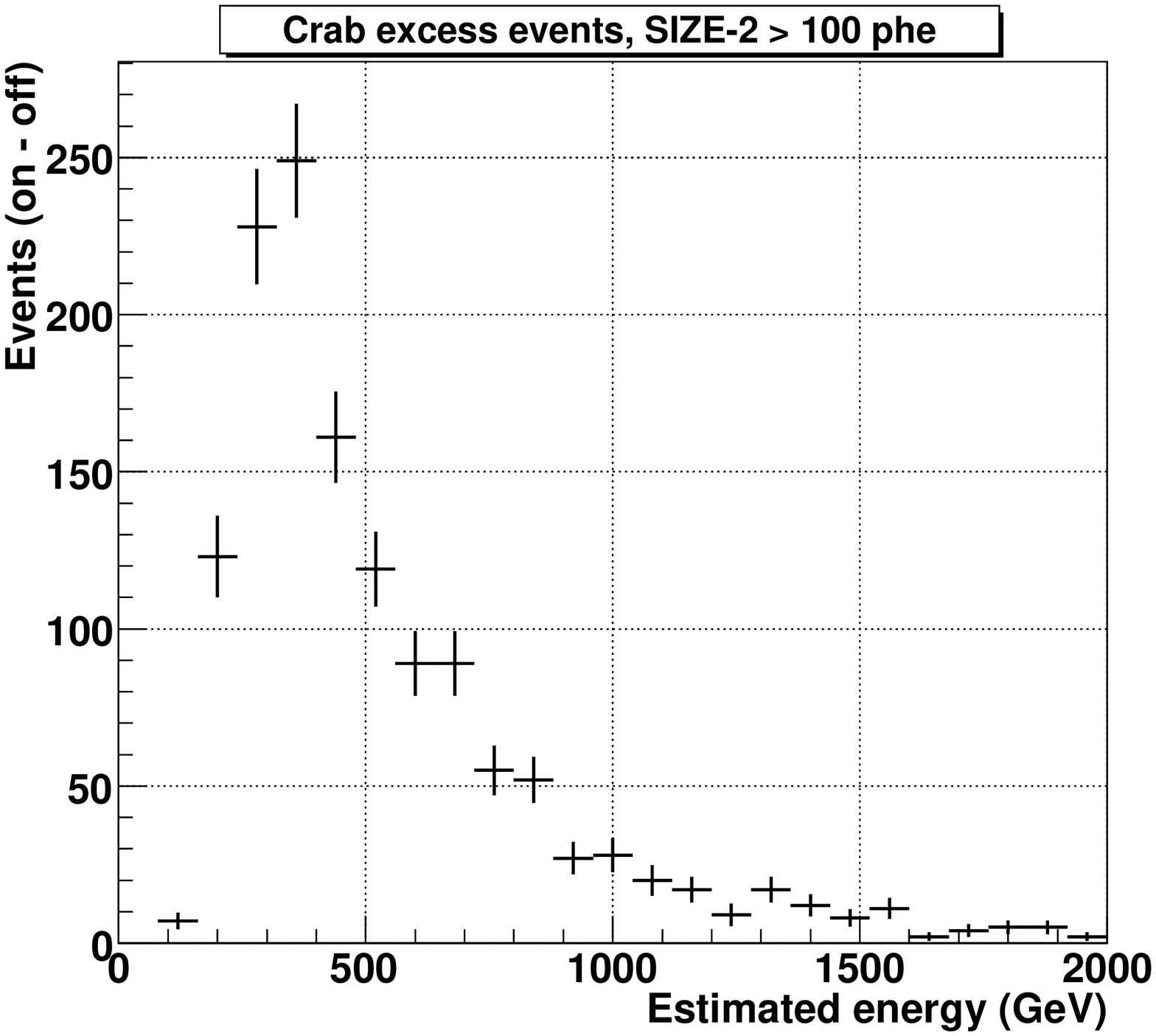}
\includegraphics[width=0.42\textwidth,height=0.38\textwidth]{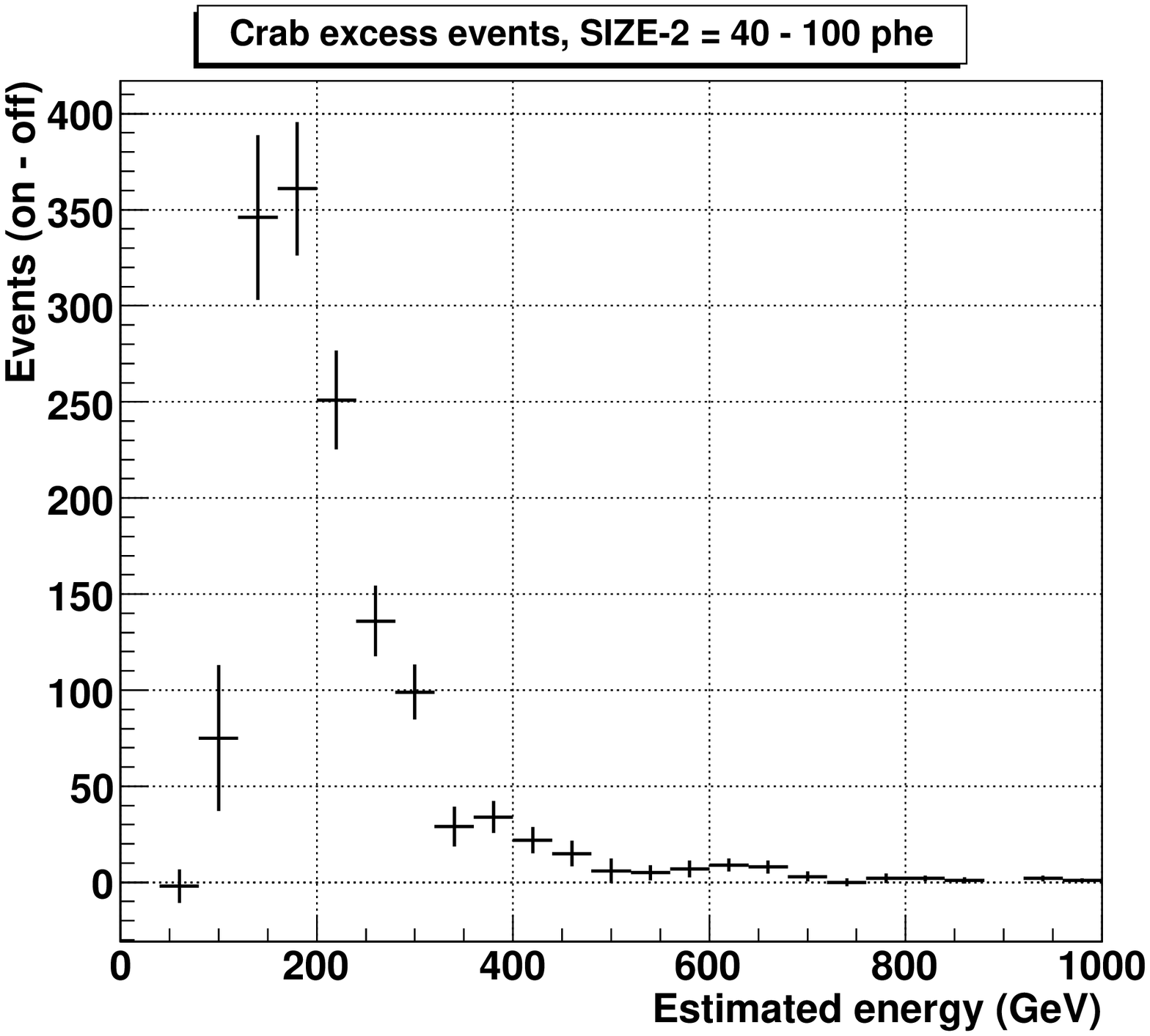}
\includegraphics[width=0.42\textwidth,height=0.38\textwidth]{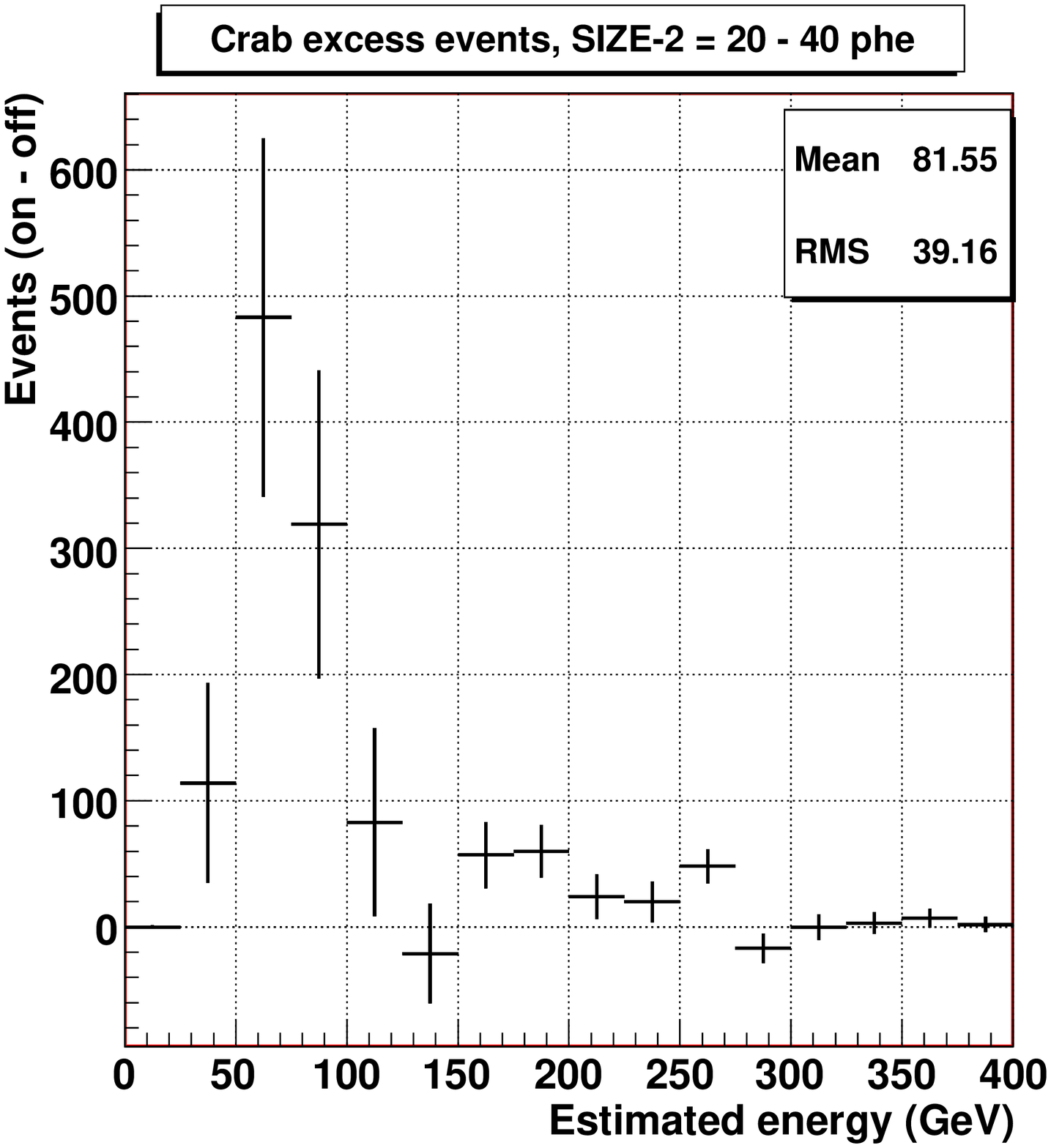}
\caption{Distribution of estimated energy for the Crab Nebula excess
  events (analysis 3) in the three ranges of \emph{Size-2} considered: above
  100 phe, from 40 to 100 phe and from 20 to 40
  phe.} \label{real_energies}
\end{center}
\end{figure}

Three different bins of \emph{Size-2} are considered in this work: the first
one (\emph{Size-2}~$>$~100~phe) corresponds to the energy range where the
integral flux sensitivity of MAGIC is best (resulting in a peak gamma
energy of around 280 GeV); the second bin
(40~phe~$<$~\emph{Size-2}~$<$~100~phe) is intended to study the performance
at intermediate energies (peak energy $\simeq 150$ GeV). Finally, the
performance for gamma rays below 100 GeV, which will be discussed in a
separate section, has been evaluated in the \emph{Size-2} range from 20 to 40
phe.
The estimated energy distributions for the excess events in each of the three \emph{Size-2} 
bins just mentioned (obtained from the real data sample) are shown in figure \ref{real_energies}.
\subsection{Background rejection} \label{sensit_comp}
\begin{figure}[!h]
\begin{center}
\includegraphics[width=0.45\textwidth, height=0.29\textwidth]{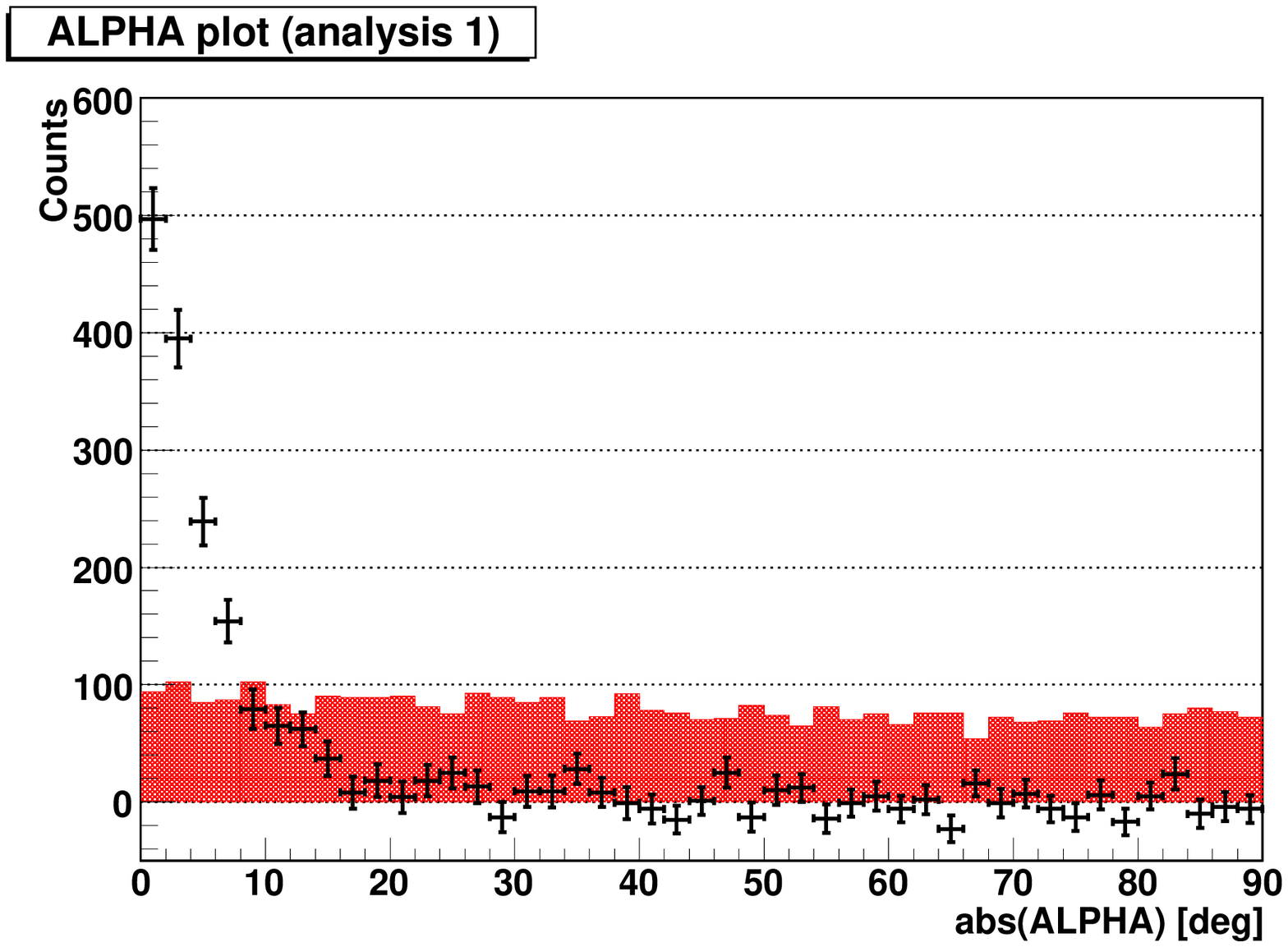}
\includegraphics[width=0.45\textwidth, height=0.29\textwidth]{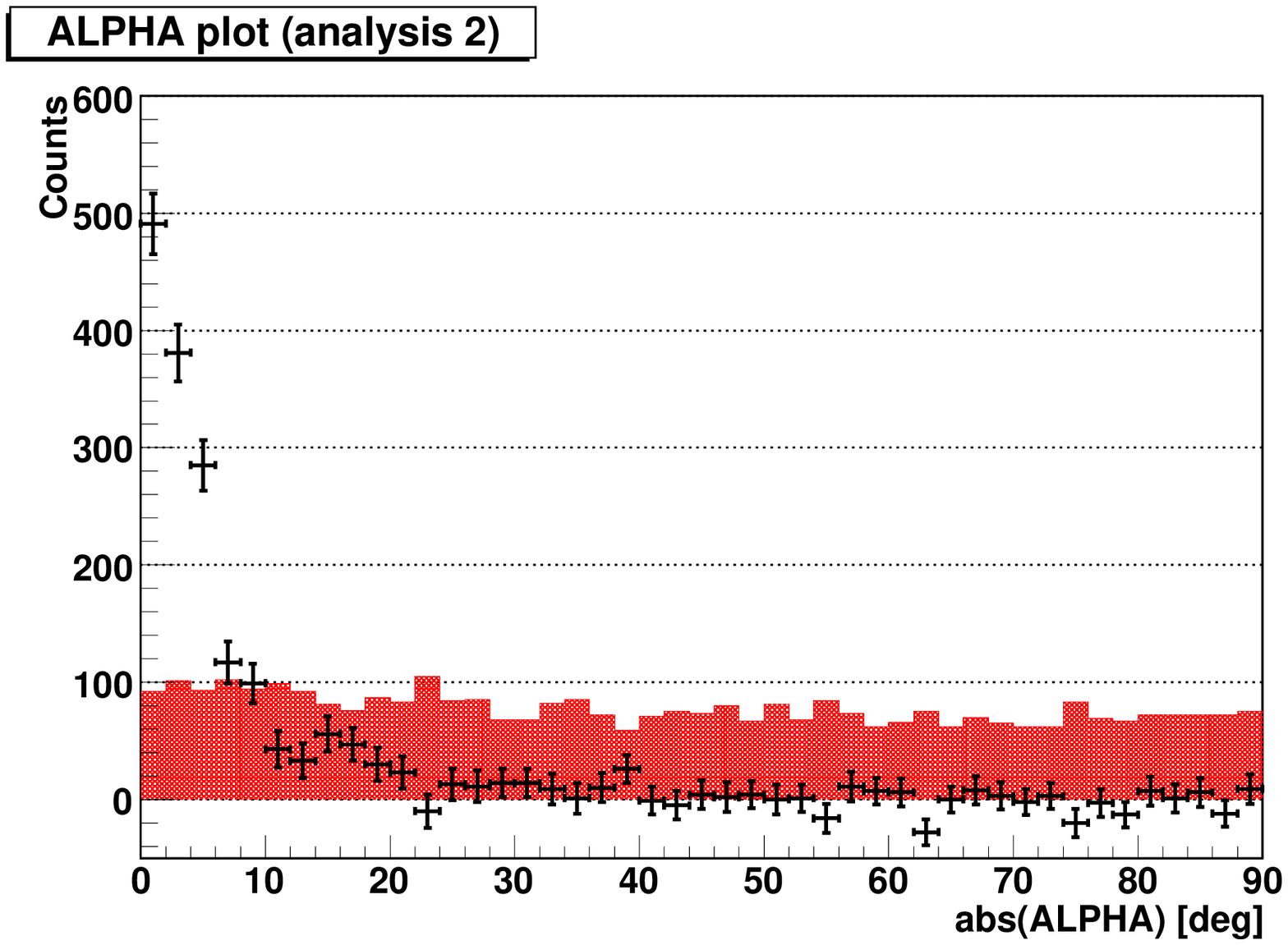}
\includegraphics[width=0.45\textwidth, height=0.29\textwidth]{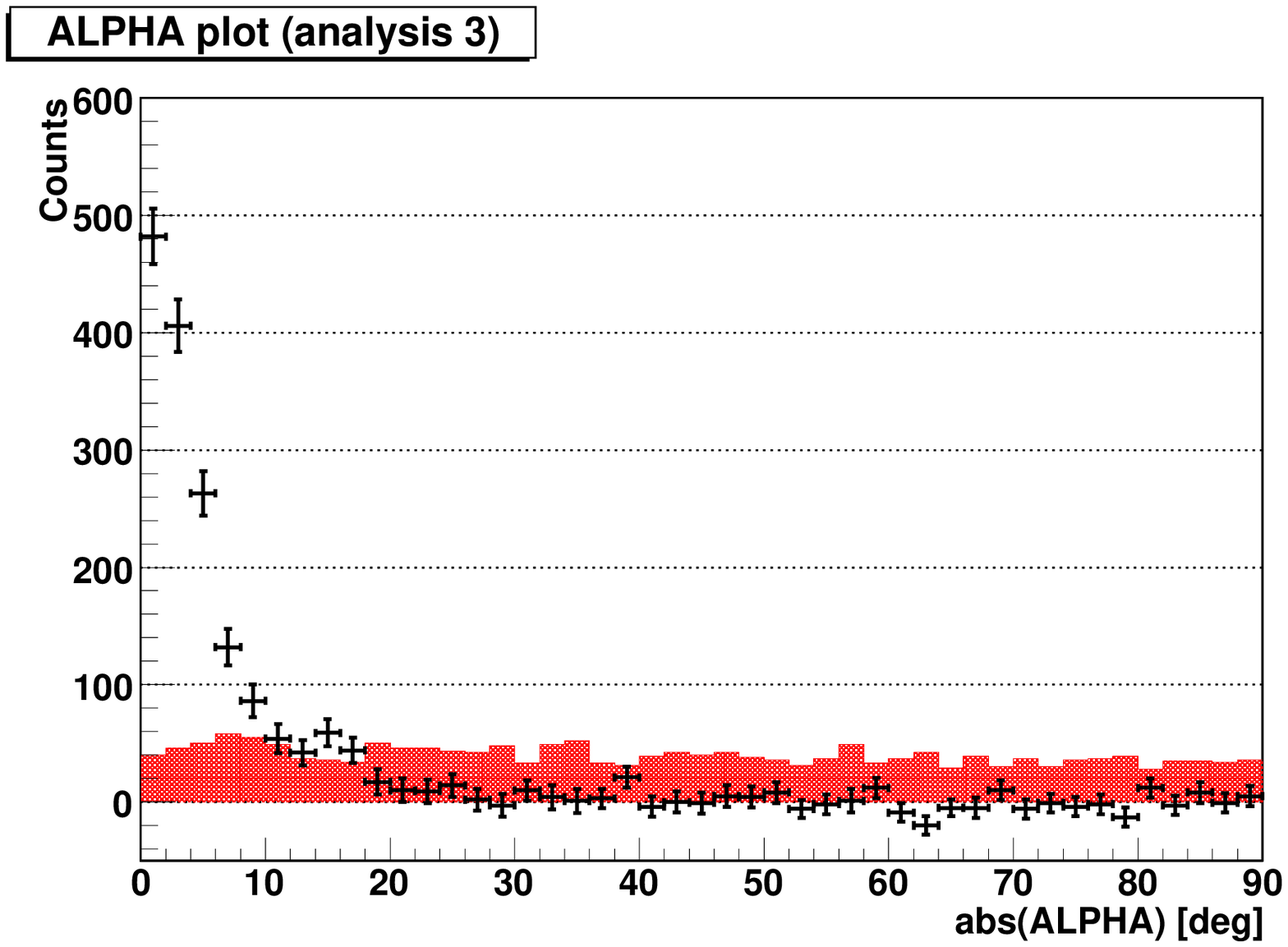}
\caption{Crab Nebula \emph{Alpha} plots (excess and residual background) obtained with
  the three tested analysis methods. 
  The \emph{Size-2} parameter is above 100~phe, corresponding to an
  energy distribution peak of $\simeq$~280 GeV. Fixed the optimal cut
  for analysis 1, the other cuts are chosen in order to have roughly
  the same number of excess events in all three
  analyses.}\label{alpha_sc100}
\end{center}
\end{figure}

\begin{figure}[!ht]
\begin{center}
\includegraphics[width=0.45\textwidth, height=0.29\textwidth]{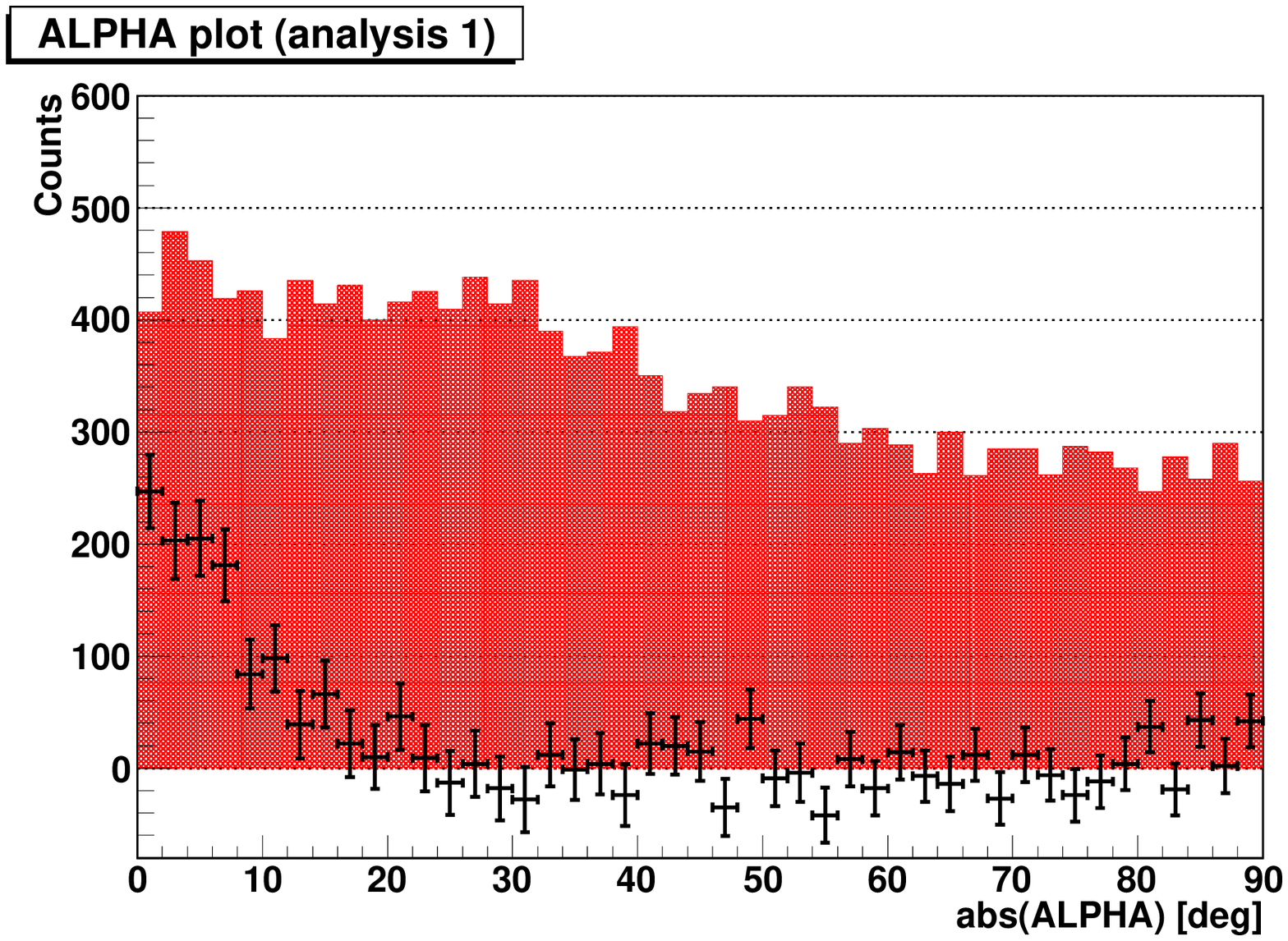}
\includegraphics[width=0.45\textwidth, height=0.29\textwidth]{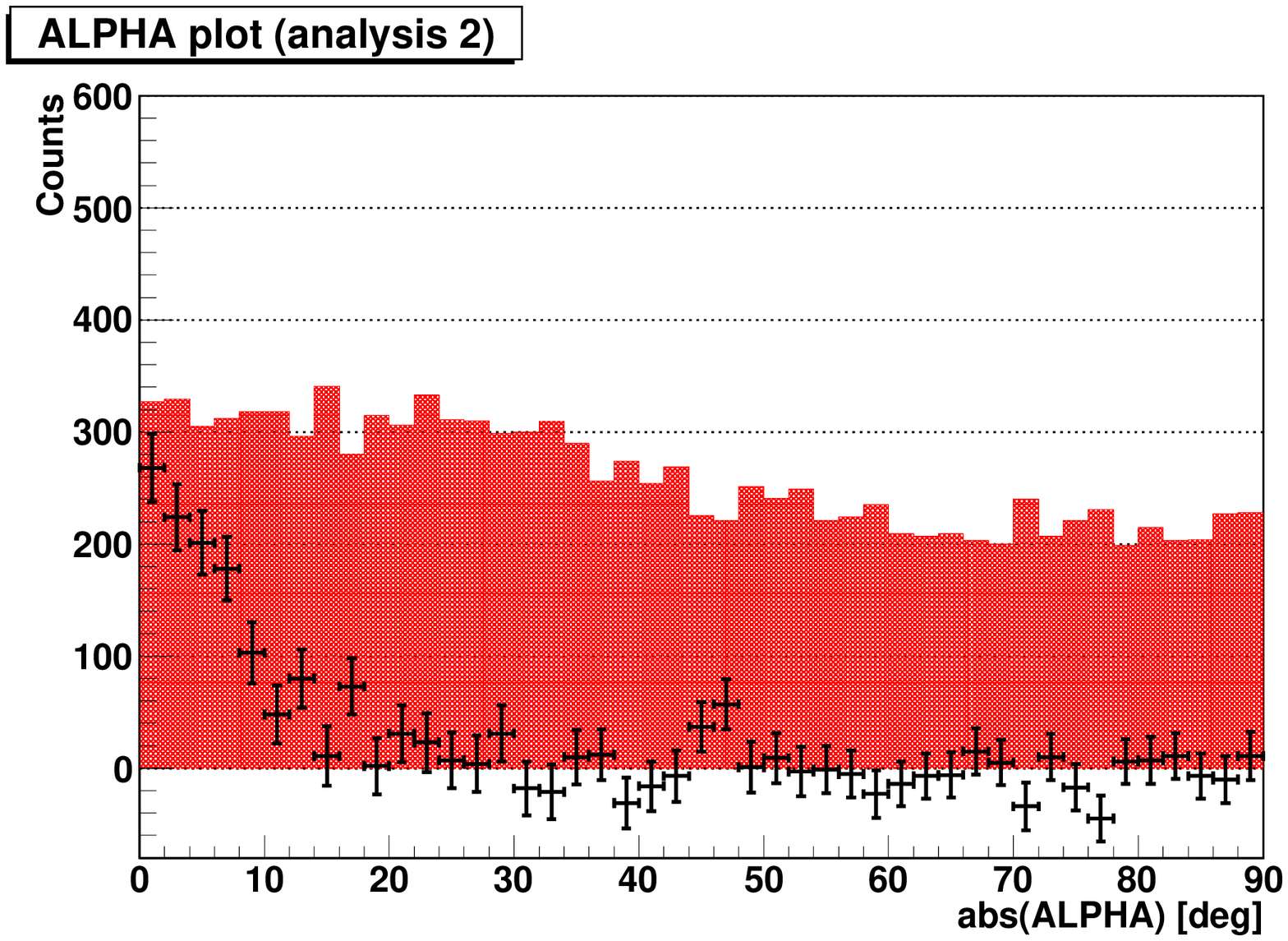}
\includegraphics[width=0.45\textwidth, height=0.29\textwidth]{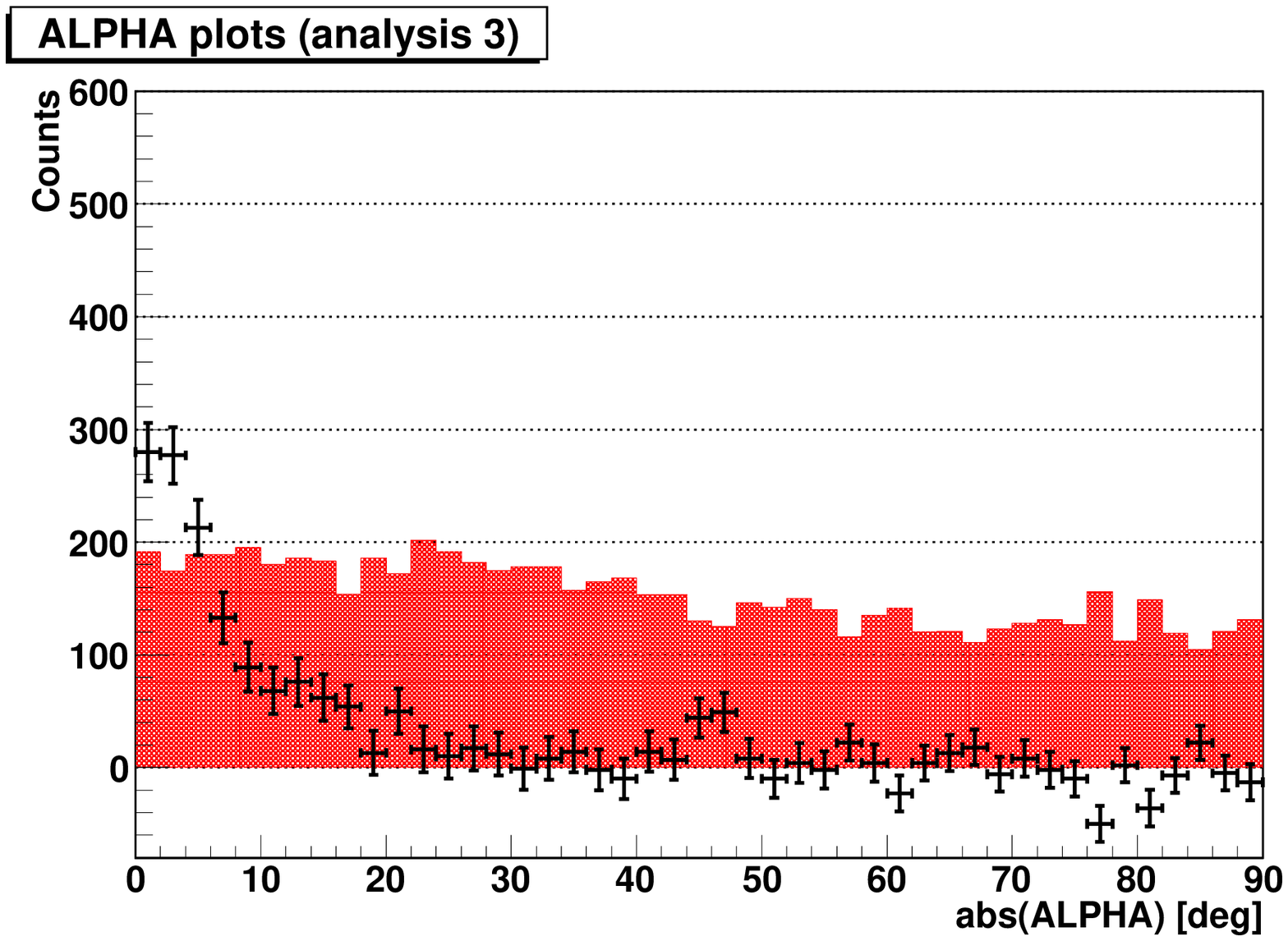}
\caption{Crab Nebula \emph{Alpha} plots (excess and residual background) from analysis 1,
  2 and 3, in the \emph{Size-2} range from 40 to 100~phe, corresponding to an
  energy distribution peak of $\simeq$150~GeV.}
\label{alpha_sc40-100}
\end{center}
\end{figure}

For the two higher \emph{Size-2} bins considered, a series of three \emph{Alpha}
plots are shown in figures \ref{alpha_sc100} and \ref{alpha_sc40-100}.
The first \emph{Alpha} plot is relative to the standard analysis
(1), the second to the time cleaning analysis (2) and the third to the time cleaning 
and time parameters (3). 
In the case of analysis 1, the \emph{Alpha} and \emph{Hadronness} cuts are optimized
to obtain the best statistical significance of the excess. 
For the analyses 2 and 3, the \emph{Hadronness} cut was chosen so that
we got the same number of excess events as in analysis 1 (after applying the same \emph{Alpha} cut). 
In this way we can easily compare the background suppression provided by each
analysis procedure. 
\begin{table}[!h]
\begin{tabular}{*{6}{c}}
\hline
Analysis & HADR. cut & \emph{Alpha} cut & Excess  & Background & $\sigma_{Li\&Ma}$/$\sqrt{h}$  \\
 & & (deg) & ($\gamma$/min) & (events/min) & \\
\hline
1  & 0.09   & 8  &3.78$\pm$0.13   & 1.08$\pm$0.06 & 12.5\\
2  & 0.10    & 8  &3.75$\pm$0.13   & 1.14$\pm$0.06 & 12.3\\
3  & 0.07   & 8  &3.78$\pm$0.12   & 0.57$\pm$0.04 & 14.0\\
\hline
\end{tabular}
\caption{Statistics of the plots in figure \ref{alpha_sc100} (\emph{Size-2}
  $>$ 100~phe; $E_{peak} \simeq 280$ GeV), obtained with 5.7 h of
  observation.} \label{tab1}
\end{table}
\begin{table}[!ht]
\begin{tabular}{*{6}{c}}
\hline
Analysis & HADR. cut & \emph{Alpha} cut & Excess        & Background & $\sigma_{Li\&Ma}/\sqrt{h}$  \\
 & & (deg) & ($\gamma$/min) & (events/min) & \\
\hline
1 & 0.10   & 12  & 3.00$\pm$0.23   & 7.58$\pm$0.15   & 5.5\\
2 & 0.09  & 12  & 3.01$\pm$0.21   & 5.62$\pm$0.13   & 6.2\\
3 & 0.07  & 12  & 3.12$\pm$0.17   & 3.29$\pm$0.10   & 7.8\\
\hline
\end{tabular}
\caption{Statistics of figure \ref{alpha_sc40-100} (40 phe $<$ \emph{Size-2}
  $<$ 100 phe; $E_{peak} \simeq 150$~GeV), obtained with 5.7 h of
  observation.} \label{tab2}
\end{table}
Note that the histograms with error bars represent the \emph{Alpha}
distribution of the excess events, instead of the usual plot showing the \emph{on}-source 
data before the background subtraction. 
In this way we can immediately see that the gamma excess is similar in all
three analyses, regardless of the background level.
\par
The main result from this comparison is that the use of the time
cleaning and the time parameters allows to halve the residual
background while keeping the same number of excess events, with
respect to the analysis using no time information. 
This can clearly be seen in figures \ref{alpha_sc100} and \ref{alpha_sc40-100} 
and the corresponding tables, Tab. \ref{tab1} and \ref{tab2}. 
Note that the quoted significance values are calculated using only 
%the anti-source
one false-source position 
for the background estimation, so the ratio of \emph{on}-source to 
\emph{off}-source exposure is one. 
The results for the lowest \emph{Size-2} bin will be discussed in section \ref{lowE}.
\subsection{Flux sensitivity to point sources} \label{sec_flux_sensi}
From the results of the Crab Nebula observations we can estimate the
flux sensitivity to point sources achievable with the different
analyses. We define the flux sensitivity as the minimum gamma ray flux
detectable in 50 hours, where ``detectable'' means that the excess of
gamma rays corresponds to a signal to noise ratio of five ({\small
  $N_{exc}/\sqrt{N_{bg}} = 5$}). 
This is the standard definition commonly used in the field, but note that 
it does not correspond exactly to a ``$5\sigma$ detection'', because the real 
significance is usually computed with the Li and Ma formula \cite{LiMa} which 
takes into account the uncertainty in the determination of the background.
\begin{figure}[!h]
\begin{center}
\includegraphics[width=0.7\textwidth]{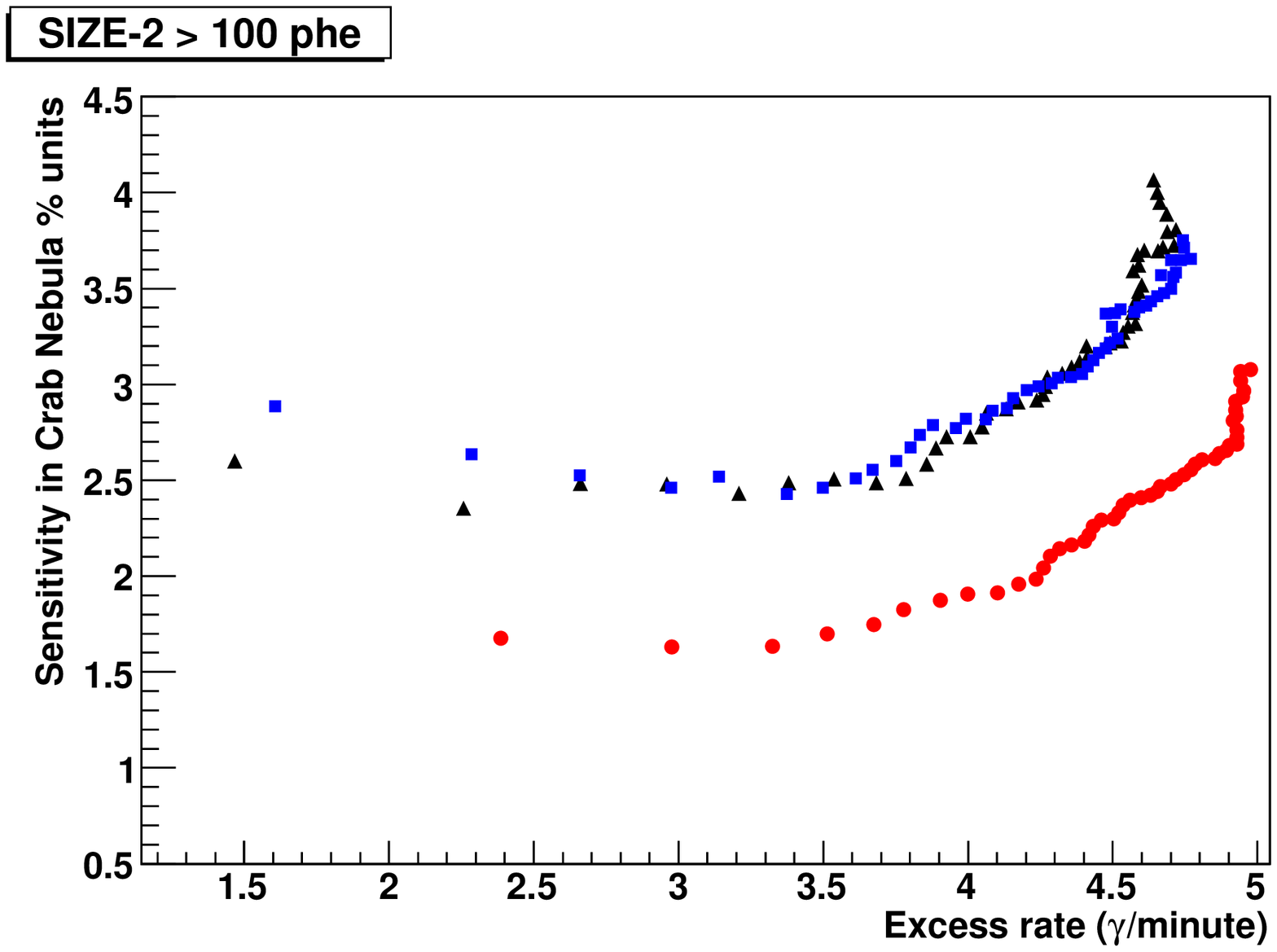}
\includegraphics[width=0.7\textwidth]{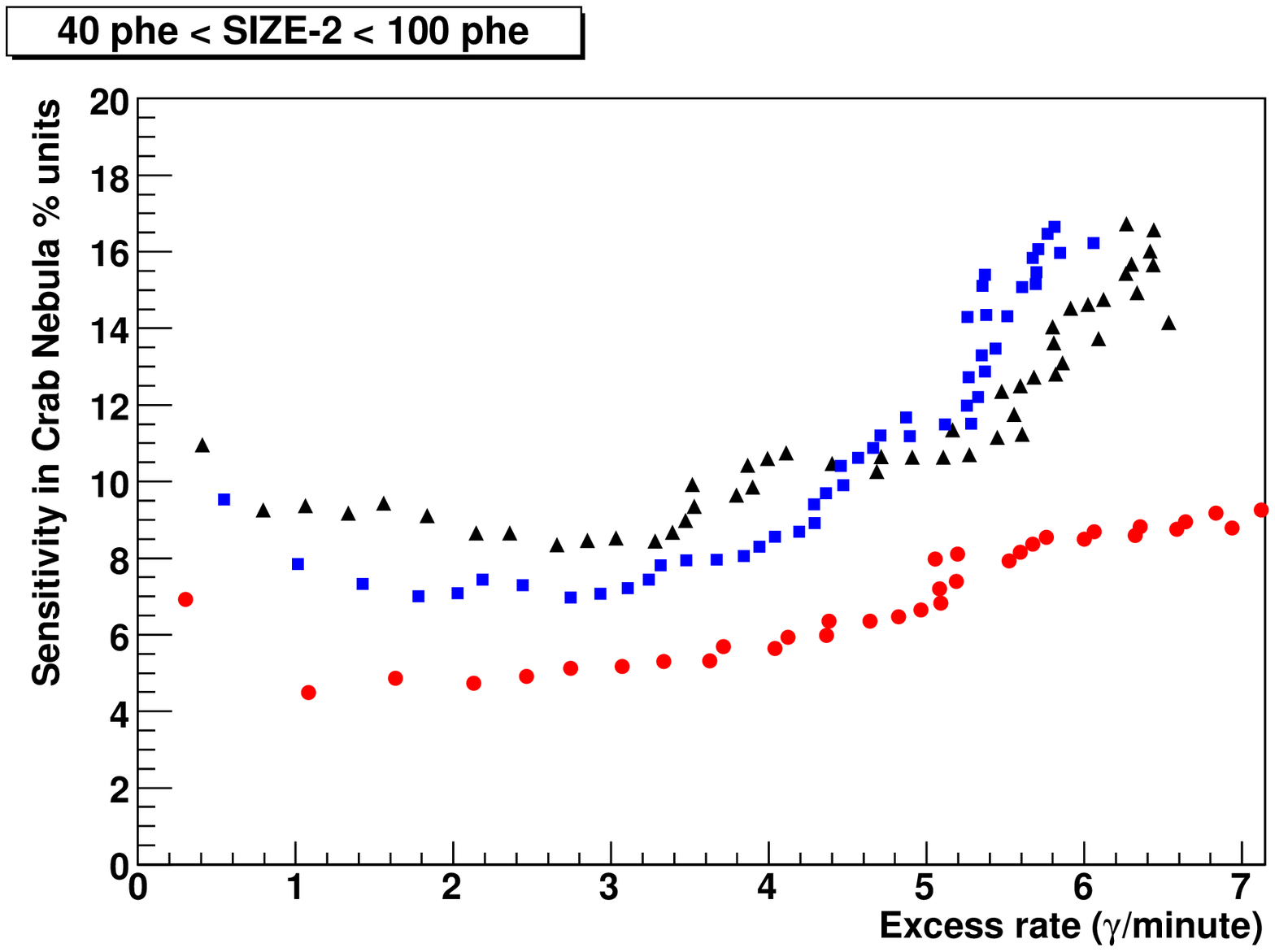}
\caption{Sensitivity (for 50 h) curves as function of the rate of
  gamma rays after cuts. The black triangles, blue squares and red circles correspond to
  analysis 1, 2,  and 3, respectively. Upper panel: \emph{Size-2}~$>$~100~phe,
  corresponding to an energy distribution peak of $\simeq$~280 GeV.
  Lower panel: 40~$<$~\emph{Size-2}~$<$~100~phe, corresponding to an energy distribution peak of $\simeq$~150~GeV. The curves are obtained by a scan of the cut in the \emph{Hadronness} parameter.} \label{sensi_both} 
\end{center}
\end{figure}
\par
The flux sensitivity depends on the strength of the background
discrimination cut (\emph{Hadronness} $< H_{max}$). 
Actually, the cuts which maximize the statistical significance of the excess from a strong
source like the Crab Nebula, as used in the previous section, are not the ones 
resulting in the best flux sensitivity: weak sources require tighter cuts. 
In this section we present the results of a scan of the \emph{Hadronness} cut values, 
shown in figure \ref{sensi_both}: the flux sensitivity (in percentage of the flux of the 
Crab Nebula) is plotted as a function of the rate of excess events.
Each \emph{Hadronness} cut of the scan leads to a different rate of excess and background 
events and thus to a different flux sensitivity.
The figures correspond to the two \emph{Size-2} bins considered in the previous section. 
The black triangles represent the standard analysis 1, whereas the blue squares and red
circles refer to the analysis 2 (with the time cleaning) and 3 (time cleaning and time-related parameters). 
Note that, since the values are derived from real Crab Nebula observations, the flux
percentage is relative to the true Crab flux, and not to the simple power-law spectrum that 
is often assumed in sensitivity estimates based on MC.
For each choice of \emph{Hadronness},  a fixed \emph{Alpha} cut (of $7^\circ$ and $10^\circ$ respectively) was applied in order to compute the sensitivity. 
The improvement coming from the use of timing in the analysis is clear in both cases. 
It must be noted that in the higher energy bin, all of the improvement comes from the 
use of the timing parameters, whereas in the lower one the introduction of the time
cleaning already results in some improvement in sensitivity. 
The best integral sensitivity that can be reached is around 1.6\% of the Crab flux for a 
peak energy of 280~GeV (left panel of figure \ref{sensi_both}).
\par
We have computed also the flux sensitivities in differential bins of estimated energy for analysis 3,
shown in table \ref{table_diffsensi}.
\begin{table}[!h]
\begin{center}
\begin{tabular}{*{7}{c}}
\hline
$E_{est}$ range & H. cut & $\alpha$ cut & Excess & Backg. & Sensitivity \\
(GeV) & & (deg) & ($\gamma$/min) & (events/min)& (\% Crab) \\
\hline
100 $<E<$ 200   &0.02   & 12  & 0.70$\pm$0.06  & 0.23$\pm$0.03 &  6.3 \\
200 $<E<$ 300   &0.02   & 8   & 0.80$\pm$0.06  & 0.11$\pm$0.02 &  3.7 \\
300 $<E<$ 500   &0.02   & 8   & 1.00$\pm$0.06  & 0.09$\pm$0.02 &  2.8 \\
500 $<E<$ 1000  &0.04   & 4   & 0.79$\pm$0.05  & 0.03$\pm$0.01 &  2.1 \\
$E>$ 1000 &0.06   & 4   & 0.28$\pm$0.03  & 0.005$\pm$0.003 &  2.1 \\
\hline
\end{tabular}
\end{center}
\caption{Sensitivity (\% Crab in 50 h) and statistics for some
  differential energy bins using the time cleaning and the timing
  parameters in the analysis 3 of section \ref{analyses}. Cuts
  are optimized separately in each bin with the best sensitivity
  criteria. Observation time: 5.7~h.} \label{table_diffsensi} 
\end{table}
\subsection{Use of timing at lower energies\label{lowE}}
The background suppression capabilities degrade as we move towards lower energies. 
This trend can be clearly seen by comparing figures \ref{alpha_sc100} and \ref{alpha_sc40-100} 
and their corresponding tables: if we focus on analysis 3, we notice that we
move from having a signal nearly seven times larger than the residual
background, to having a signal (integrated below the \emph{Alpha} cut)
slightly smaller than the background. 
This is mainly a result of the worsening of the gamma/hadron discrimination and of the 
fact that the spectrum of the Crab Nebula is harder than that of the background, although 
this latter contribution is smaller.
\par
\begin{figure}[h]
\begin{center}
\includegraphics[width=0.45\textwidth,height=0.3\textwidth]{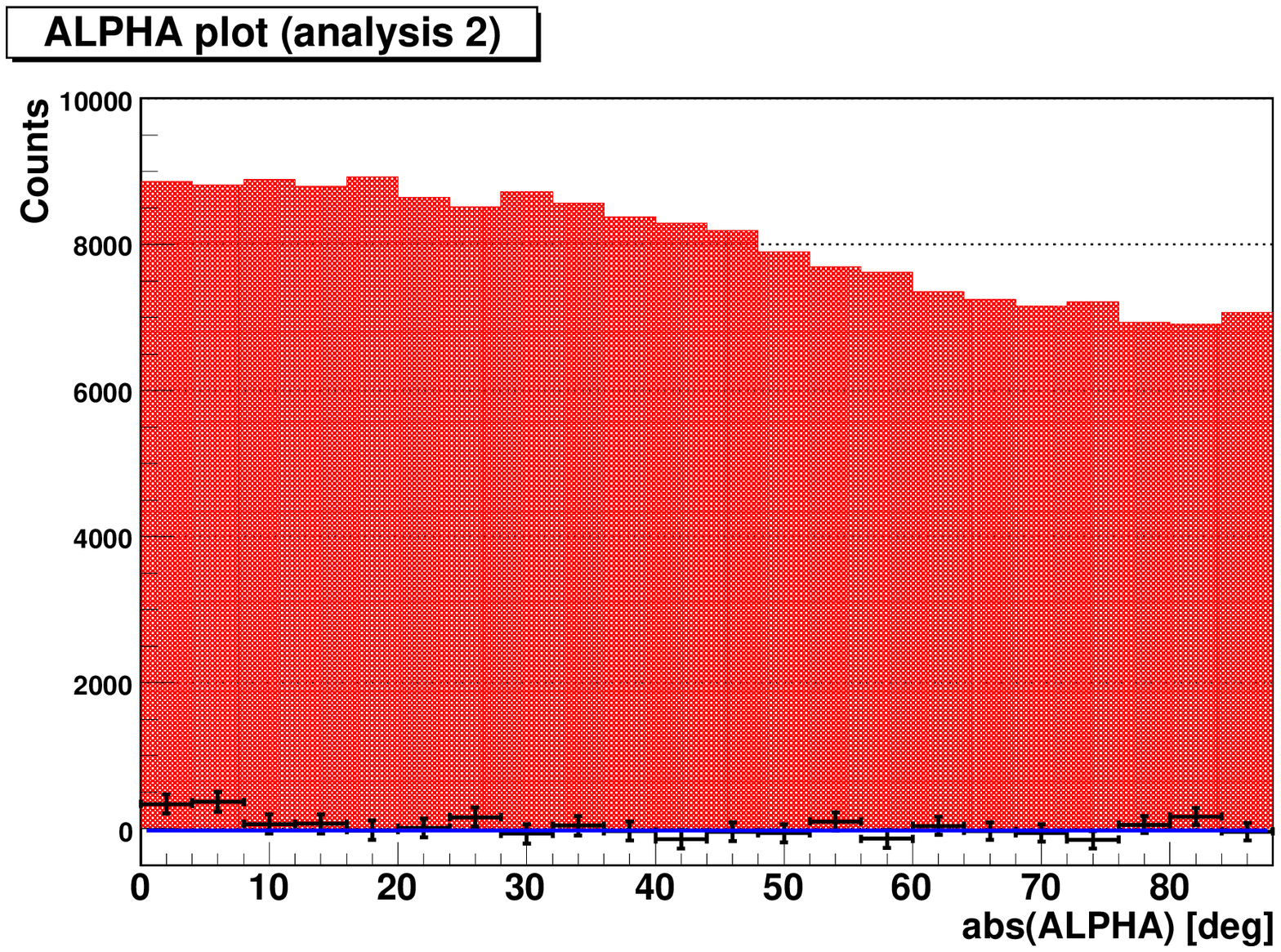}
\includegraphics[width=0.45\textwidth,height=0.3\textwidth]{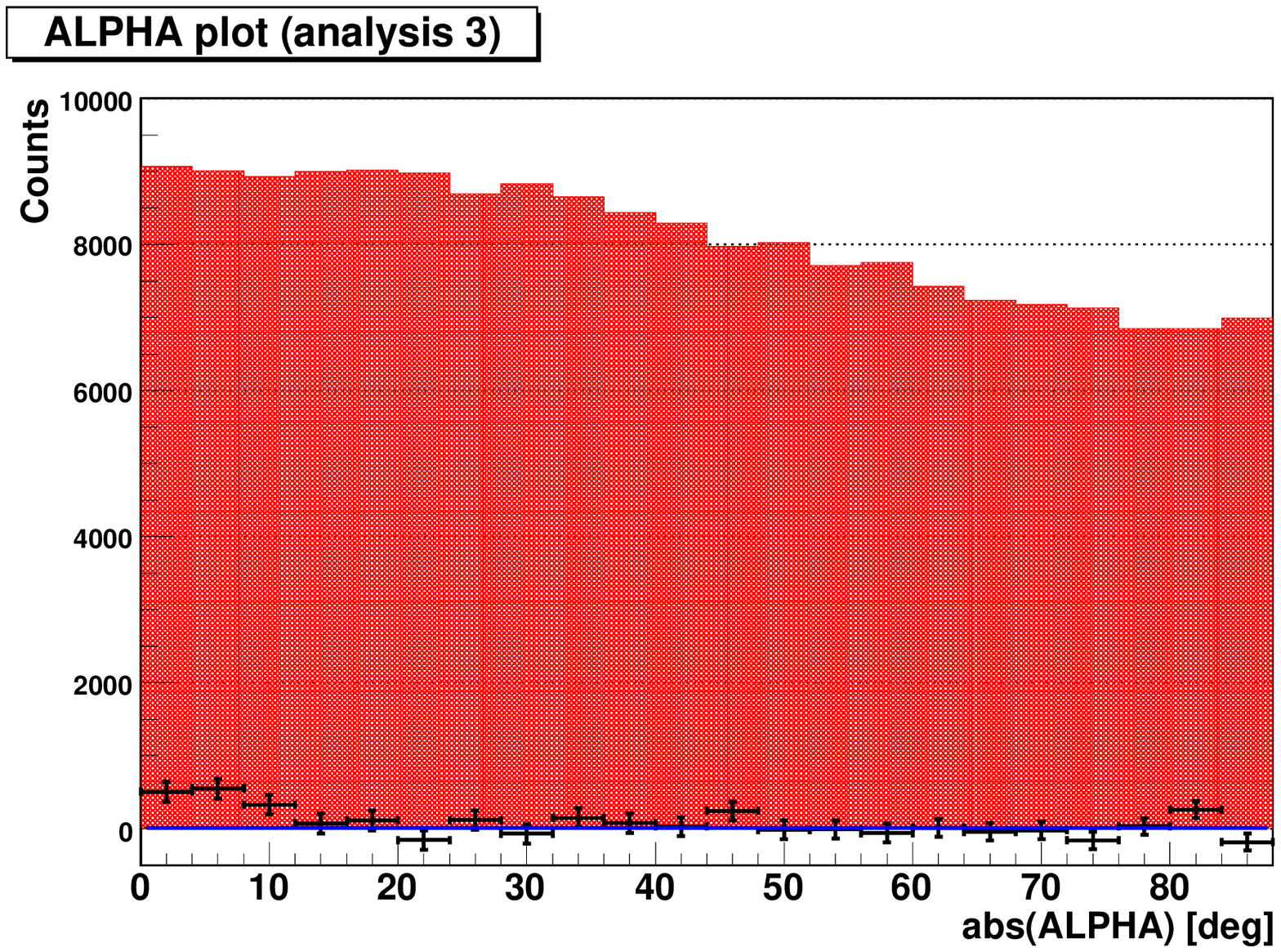}
\caption{Crab Nebula \emph{Alpha} plots (excess and residual background) from analysis
  2 and 3 in the bin of \emph{Size-2} from 20 to 40 phe (estimated energy of
  the gammas mostly below 100~GeV). Note that in this case we have
  adjusted the background rate instead of the excess rate to be the same
  in both cases. The \emph{Alpha} cut applied is of $12\deg$.} \label{alpha_low} 
\end{center}
\end{figure}

In figure \ref{alpha_low} we show the results for analyses 2 and 3 in the \emph{Size-2} 
range from 20 to 40 phe, where most of the excess comes from sub-100 GeV gamma 
rays (see third pad of figure \ref{real_energies}). 
Given the modest signal (a mere 5.7 $\sigma$ significance in analysis 3), we have 
in this case adjusted the cuts to obtain the same background rate ($\simeq 80$
events/min) in both analyses, and then compared the gamma ray excesses. 
Once more, the improvement in performance due to the introduction of the timing is clear, 
though less significant due to the large residual background. 
With roughly the same background rate the excess rate for analysis 2 is 2.5$\pm$0.7
$\gamma$/min whereas for analysis 3 it is 4.0$\pm$0.7 $\gamma$/min. 
In this energy range, analysis 1 even fails to produce a significant signal, due to the high cleaning 
levels.
\par
%For even lower energies, those of our third bin of \emph{Size-2} = 20 to 40 photoelectrons, 
For these low energies,
the background overwhelms the signal even of
a strong source like Crab, and is actually setting a further
limitation for the observation of weak sources.
The signal must not only be statistically significant, but also well above 
the systematic uncertainty in the determination of the background, which 
is at least of a few percent and unlike statistical significance, it does not
get better with longer observation times. \\
%The degradation of the background discrimination can be attributed to the smaller size of the images, approaching to the size of the camera pixels, and also to the nature of the background that comes into play, for instance distant muons making gamma-like images \cite{2006APh....25..342M} or single electromagnetic subshowers coming from hadronic primaries. 
%A detailed discussion of this issue is beyond the scope of this paper, but we will show in this section that the use of timing information is also helpful in the low energy regime, below 100 GeV.
% NEW
It has to be noted that, even after strong background rejection cuts, a certain amount of 
``irreducible'' background survives. 
The background rejection power of the imaging technique degrades fast with the lowering 
of the energy of the primary gamma-ray. 
The reason of this degradation can be attributed both to the physics of the air showers 
and to the technical limitations of the IACT instruments. 
The irreducible background is made up by shower images which are similar to gamma 
induced images in all of the image parameters used for the event discrimination. 
This may reflect the shortcomings of the instrument in recording the small and faint images 
of low energy showers, resulting from its limited light collection efficiency and camera pixelization. 
A larger reflecting surface or an increased quantum efficiency camera, together with a finer 
pixelization, would obviously improve the accuracy of the reconstructed image parameters. 
On the other hand, even assuming a perfect IACT detector, there is no guarantee that the intrinsic characteristics of the cascades are different enough to permit to distinguish the nature of the 
primary particles. 
The study of the characteristics of the gamma-like background (see for example \cite{maier} or 
\cite{dorota}) heavily rely on the MC simulation packages.
The nature of this irreducible background is attributed to $\pi^0$-s, primary electrons and 
long flying relativistic particles (like $\mu$-s). 
Proton-induced air showers typically produce pions in the first interaction stage. 
The charged pions decay into muons, whereas the $\pi^0$-s (most often decaying into two gammas)
originate electromagnetic sub-showers. If the energy of the primary particle is above a few hundred 
GeV, the superposition of the sub-showers from different $\pi^0$-s with diverging trajectories 
make hadronic showers to appear wider and more ``patchy'' than gamma-initiated ones. 
As is well known, this is the feature of hadron-initiated cascades which allows best to suppress
them in the analysis of IACT data.  
However, as the energy of the hadronic primary goes down, the pion multiplicity drops, and the 
chances that the light of a single electromagnetic subshower dominates the image recorded by 
an IACT get larger. % (see \cite{dorota}). 
This means that, regardless of the characteristics of the telescope, the amount of irreducible 
hadronic background will necessarily increase as we go down in energy.
\par
Also the background images due to distant muons can easily get confused, by a single-dish 
IACT, with low energy gammas. 
Those images have generally a small size which make it difficult to recover information from 
the shower shape. 
The time spread of pixel signals has been proposed in \cite{2006APh....25..342M} as a parameter
which may be used to suppress such background, but as we will discuss in section \ref{disc}, the 
method does not seem to work efficiently for MAGIC. 
The improvement in sensitivity in the $<100$~GeV energy range is due to the contribution of the 
\emph{Time Gradient} parameter.
\par
In summary, even though the overall background discrimination worsens very fast with decreasing energy, the use of the timing parameters (mainly the \emph{Time Gradient}) has been shown to 
improve background suppression efficiently in the whole energy range of MAGIC, even below 100~GeV.
% END NEW 
%
\subsection{Application of timing analysis to older MAGIC data}
In the previous sections we have discussed exclusively MAGIC data
taken after the installation of the fast readout in February 2007. 
In earlier MAGIC data, the signal sampling was made with FADCs with 300 MSample/s. 
Before digitization, the pulses were stretched up to 6 ns FWHM
(10 ns for the low gain), to ensure proper sampling. Timing
information was only used in the image cleaning stage of the analysis,
to reduce the energy threshold \cite{2007CrabMAGIC}. Only shape
parameters were used for background suppression. With that sort of
``classical" analysis, no improvement in performance could be seen 
after the upgrade of the DAQ system: both for the data before and
after the upgrade the best integral flux sensitivity achieved in
wobble mode was around $2.4\%$ of the Crab Nebula flux in 50 hours.
The reduction of the signal integration time (and subsequent reduction
of the integrated NSB noise) does not seem to result, by itself, in an
improved performance. The reason is that the intrinsic fluctuations of
the Cherenkov light recorded by a pixel (coming from the Poissonian
photon statistics) dominate over the fluctuations of the NSB light
(with a mean rate of about 0.13 phe per nanosecond in an inner pixel),
and therefore the reduction of the NSB noise does not change
significantly the precision of the charge measurement in a
pixel\footnote{This applies to the ``dark night''  observations
discussed in this paper. For observations under 
  moonlight or in twilight, with a higher rate of NSB photons, the
  shorter integration time is indeed an advantage in the
  reconstruction of small showers.}. 
\par
Despite the poorer quality of the timing information recorded in data
taken before February 2007, we have also tried to apply the timing
parameters defined in section \ref{timepar_description} to their
analysis. A 8.7 hour Crab data sample taken in good weather
conditions during four nights in December 2006 and January 2007, shortly
before the change of the DAQ system, was used for this test. It turns
out that the improvement in background suppression brought about by the
timing parameters on these data is smaller than the one obtained on
newer data, shown in sections \ref{sensit_comp} and \ref{sec_flux_sensi}.
The best integral sensitivity, achieved for the SIZE-2$>100$ phe sample, is $1.9\%$ of
the Crab flux.
\par
Although we have not yet been able to investigate this issue in
detail, the most obvious reason for the smaller effect of the timing
parameters in the analysis of MAGIC data taken before the upgrade of
the readout is the worse quality of the timing
information. Nevertheless, the precision of the determination of the
arrival time of calibration 
pulses, estimated by studying the reconstructed times for different
pixels as discussed in section \ref{MCSimulation}, is roughly the same
for both setups ($\simeq 390$ ps RMS). Calibration flashes make rather
large pulses of around 35 photoelectrons per inner pixel
\cite{2007ICRCGoebel}, and certainly most of the pixels in the
processed shower images have smaller signals. On top of that,
calibration pulses are wider than pulses from showers (4.6 vs. 2.3
ns FWHM). That is, calibration pulses are not representative of the bulk of
the pulses which contribute to the images and thus to the time
parameters we use in the background discrimination. In conclusion, the
study mentioned above, using calibration pulses, is not in
contradiction with the na{\"{\i}}ve expectation of an improvement in the
accuracy with which pulse times are reconstructed with the faster
sampling. The better performance of timing parameters in the
suppression of the background after the FADC upgrade is an indirect
evidence that this is the case.
\subsection{Energy estimation}
The Random Forest method can also be used for the estimation of a continuous variable. 
It is the standard method used in the analysis of MAGIC data for estimating the energy 
of the showers (under the assumption that they are gamma rays). 
In an analogous way as for the background rejection, the RF is trained with MC 
gammas, whose true energy is known. 
The main difference is in the way the RF is built: the optimal cut in each node of the trees 
is chosen to minimize the variance of the true energies of the event samples resulting 
from the split \cite{RFreference}, rather than their purity. 
No background event sample is needed for this kind of training.
\par
The set of image parameters typically used for the estimation of the
energy is: \emph{Size}, \emph{Width}, \emph{Length}, \emph{Dist}, \emph{Conc}, \emph{Leakage} and \emph{Zenith Angle}. 
In order to evaluate the improvement in the energy reconstruction due
to the use of the signal timing, the energy of a test sample of
Monte~Carlo $\gamma$ events (different from the sample used for the
training) was evaluated and compared to the true known energy of the
primary gamma rays.
An improvement in the energy reconstruction is expected if the image
parameter \emph{Time Gradient} is added to the default set of parameters
since it provides information about the impact parameter of the shower. 
This should help to avoid the degeneracy between small, nearby showers 
and the large, distant ones. 
As introduced in section \ref{timepar_description}, both \emph{Dist} and \emph{Time Gradient} parameters are well correlated with the IP and can be used for its estimation. 
Notice that the correlation of \emph{Dist} is rather good for small IP event whereas for larger 
IP the estimation with \emph{Dist} becomes poorer (figure \ref{correlations1}). 
In case of \emph{Time~Gradient} the correlation is better for
higher IP values, very likely because for distant showers the time
structure of the images is more pronounced and as a consequence more
precisely measured.
\begin{figure}[!h]
\centering
\includegraphics[width=0.7\textwidth,height=0.5\textwidth]{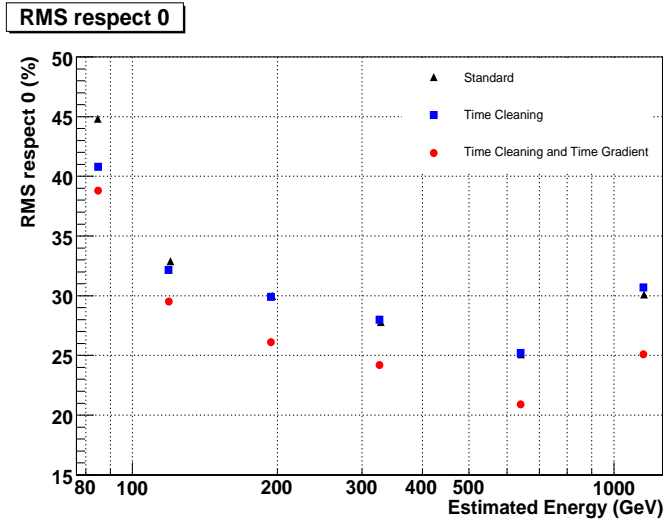}
\caption{Graph of the value of  the RMS with respect to zero of the quantity $(E_{est}-E_{true})/E_{est}$ for different energy bins. This value is an overall estimator of the energy resolution. The black triangle and the blue square show data processed with the standard set of energy estimation parameters (image cleaning 10-5 phe and time image cleaning 6-3 phe respectively). The red circles correspond to data processed with time image cleaning and also \emph{Time Gradient} used as energy estimation parameter (time image cleaning 6-3 phe). 
} \label{en_res}
\end{figure}
\par
The distributions of the quantity $(E_{rec}-E_{true})/E_{rec}$ for different energy bins
were used to estimate the quality of the energy reconstruction.
In fig. \ref{en_res} the black triangles refer to the energy estimation with the standard 
parameter set, that is, data processed with the 10-5~phe image cleaning without time constraints. 
The blue squares correspond to an energy reconstruction performed with the 
standard parameter set and the time image cleaning 6-3~phe (see section \ref{cleaning}), 
while the red circles are obtained from the time-constrained
image cleaning and the \emph{Time Gradient} image parameter being added to the  
standard set for the energy estimation.
The graph represents the value of the RMS of $(E_{est}-E_{true})/E_{est}$ with respect 
to zero instead of the mean value. 
This quantity is preferred to the simple RMS as an overall estimator of the quality 
of the energy reconstruction, since it takes into account not only the spread of the distribution, 
but also a possible bias with respect to zero (see \cite{tescaro_tesina} for more details).
``Leakage" effects for images located close to the edge of the camera could be 
important when the energy reconstruction is performed since the number of photons 
in the part of the image outside the camera is actually not measured. 
A standard selection cut \emph{Leakage}~$<$~10\% is applied in this analysis
and tests with tighter \emph{Leakage} cuts revealed no significant changes
with respect to the shown results. 
The use of the \emph{Time Gradient} image parameter in the energy estimation yields 
to a relative improvement in energy reconstruction of around 15\%. 
%
%%%%%%%%%%%%%%%%% DISCUSSION %%%%%%%%%%%%%%%%%%%%
%
\section{Discussion}\label{disc}
We have established that a significant improvement of performance of the MAGIC 
Cherenkov telescope, both in terms of flux sensitivity and energy resolution, 
can be achieved by using the timing of signals in the reconstruction of shower images.
The reason for the improvement is two-fold: on one hand, the time-constrained image 
cleaning allows to reduce the cleaning charge levels without adding noise coming mainly 
from the night sky background light, which results in a lower analysis energy threshold. 
On the other hand, the timing profile of the images, represented by the \emph{Time Gradient}, 
provides information about the shower impact parameter, a relevant quantity which is 
otherwise poorly determined by a single-dish IACT.\\
The results presented here have been reproduced also by a different analysis within the MAGIC collaboration. This analysis shares the ``core'' of MARS and is based on \cite{BretzRiegel} but the algorithms for the analysis are developed independently, for example fixed cuts in combinations of the image parameters 
are used instead of the Random Forest.
Also in this case is the introduction of the \emph{Time Gradient} led to the described improvement in 
the background suppression.
\begin{figure}[b!]
   \centering
   \includegraphics[width=0.75\textwidth,
   height=0.55\textwidth]{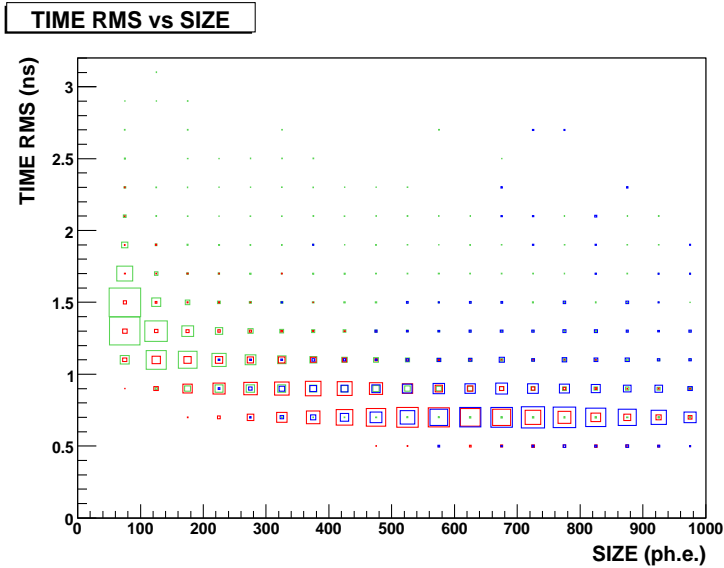} 
   \caption{\emph{Time RMS} versus \emph{Size} of the event. The green color refers
     to simulated gammas, red to simulated muons and blue to clear
     real muon events (an image arc of at least 180$^\circ$ is
     required). 
     Gammas and muons extensively overlapped in the considered parameters space. 
     Notice the absence of events below
     500~ps, evidence of our blindness to time features below this threshold,
     and the increase of the time RMS values at very low sizes
     likely due to the worsening of the time resolution for very small signals.} \label{muons}
\end{figure}
\par
From the sensitivity graphs of section \ref{sensit_comp}, it is possible to conclude that the 
time cleaning alone results in a significant sensitivity improvement in the low energy regime
(40~$<$~\emph{Size-2}~$<$~100 phe), coming from the increased event statistics. 
At higher energies it does neither improve nor worsen the telescope performance 
significantly although the lower cleaning results in more pixels per image.
In contrast, the \emph{Time Gradient} seems to be helpful in the entire energy range 
accessible to MAGIC. 
This parameter allows to reject hadron showers whose images are gamma-like in shape 
and oriented towards the gamma-ray source location on the camera, but whose \emph{Dist} 
and \emph{Time Gradient} parameters are not consistent with what is expected for a gamma 
shower coming from a point source.
\par
The events with very large \emph{Time~RMS} ($\gtrsim$~1.5~ns) are rejected thanks to
the \emph{Time RMS} parameter.
The background rejection power shown by the \emph{Time~RMS} in this study is much lower 
than foreseen in \cite{2006APh....25..342M}. 
This is most likely due to the too optimistic assumptions regarding the telescope features 
made by the authors of that work with respect to the actual characteristics of MAGIC, in particular 
regarding the reflecting dish. 
In the final mounting the panels of the MAGIC mirror are staggered in a chessboard 
pattern to facilitate their movement and to ensure a proper focusing, and this  causes the 
parabolic dish not to be perfectly synchronous.
The mirrors staggering together with the other sources of time spread in the acquisition chain, like the jitter in the transit time of the electrons in the PMTs, lead to a time-RMS response larger than expected
from the authors of \cite{2006APh....25..342M}. 
The value that results from almost synchronous input signals, for example muon events, 
is $\approx$0.7~ns, a value comparable with the intrinsic time spread of the low-energy 
$\gamma$-events (see figure \ref{muons}). 
Therefore, the tagging of single distant muons from just their time spread is at the 
moment not possible.
\par
In conclusion the use of timing information in the analysis of MAGIC data provides a
considerably better background suppression and results in an enhancement of about 
a factor 1.4 of the flux sensitivity to point-like sources, as tested on real observations of 
the Crab Nebula. 
This gain is equivalent to doubling the available observation time.\\ 
Improvements of the order of 15\% have been found in the event energy reconstruction. 
In fact the time gradient gives information about the real impact parameter of the shower 
and therefore it helps to distinguish distant high energy showers from closer, low energy
ones.\\ 
We expect that this type of timing analysis may also be helpful to future Cherenkov telescopes.
Even if the \emph{Time Gradient} is very likely not useful for stereo IACT systems, this does not exclude that different time-related image parameters can be worth for the reduction of the stereo system data.
The time image cleaning algorithm would be instead worth for either stereo and single IACT systems.\\ 
\section{Acknowledgements}
We would like to thank the Instituto de Astrofisica de 
Canarias for the excellent working conditions at the 
Observatorio del Roque de los Muchachos in La Palma. 
The support of the German BMBF and MPG, the Italian INFN 
and Spanish MCINN is gratefully acknowledged. 
This work was also supported by ETH Research Grant 
TH 34/043, by the Polish MNiSzW Grant N N203 390834, 
and by the YIP of the Helmholtz Gemeinschaft.
%

% \appendix
% \section{}
% \label{}

%

%%
\end{document}